%% file: main.tex
\documentclass[11pt,authoryear,a4paper]{elsarticle}

\usepackage[utf8]{inputenc}
\usepackage[hidelinks]{hyperref}

\usepackage{url}
\usepackage{xurl}
\usepackage{caption}
\usepackage{subcaption}
\usepackage{siunitx}
\usepackage{amssymb}
\usepackage{pifont}
\usepackage{makecell}
\usepackage{xspace}
\usepackage{natbib}
\usepackage{svg}
\usepackage{booktabs}
\usepackage{tabularx}
\usepackage{array}
\usepackage{multirow}
\usepackage{xcolor}
\usepackage{colortbl}

\usepackage{tikz}
\usetikzlibrary{positioning}

\usepackage{threeparttable}

\usepackage{todonotes}
\usepackage{soul} 
\soulregister\cite7 
\soulregister\ref7 

\usepackage{lineno}

\usepackage{geometry}
\usepackage{lipsum} 
\usepackage{enumerate}

\newcommand{\VASPA}{VASP-2\xspace}
\newcommand{\VASPB}{VASP-12\xspace}
\newcommand{\VASPC}{VASP-9\xspace}
\newcommand{\VASPD}{VASP-5\xspace}

\colorlet{mycol}{red}

\makeatletter
\def\ps@pprintTitle{%
	\let\@oddhead\@empty
	\let\@evenhead\@empty
	\let\@evenfoot\@oddfoot
}
\makeatother

\begin{document}

\begin{frontmatter}


\title{Assessing the Solvency of Virtual Asset Service Providers: \\Are Current Standards Sufficient?}



\author{Pietro Saggese\fnref{fn1}}
\ead{saggese@csh.ac.at}

\author{Esther Segalla\fnref{fn2}}
\ead{esther.segalla@oenb.at}

\author{Michael Sigmund\fnref{fn2}}
\ead{michael.sigmund@oenb.at}

\author{Burkhard Raunig\fnref{fn2}}
\ead{burkhard.raunig@oenb.at}

\author{Felix Zangerl\fnref{fn3}}
\ead{felix.zangerl@fma.gv.at}

\author{Bernhard Haslhofer\fnref{fn4}}
\ead{haslhofer@csh.ac.at}

\fntext[fn1]{Complexity Science Hub Vienna (CSH), Josefstaedter Strasse 39, 1080 Vienna, Austria and AIT - Austrian Institute of Technology, Giefinggasse 4, 1210 Vienna, Austria. \textit{Corresponding author.}}
\fntext[fn2]{Oesterreichische Nationalbank (OeNB), Otto-Wagner-Platz 3, 1090 Vienna, Austria.}
\fntext[fn3]{Austrian Financial Market Authority (FMA), Otto-Wagner-Platz 5, 1090, Vienna, Austria.}
\fntext[fn4]{Complexity Science Hub Vienna (CSH), Josefstaedter Strasse 39, 1080 Vienna, Austria.}

\date{\today}

\input{sections_original/0_abstract}

\end{frontmatter}


\newpage

\setcounter{footnote}{0} 


\input{sections_original/1_introduction}

\input{sections_original/2_1_background}
\input{sections_original/2_2_trad_fin}

\input{sections_original/2_3_literature}
\input{sections_original/3_1_VASPs}

\input{sections_original/3_2_comparison}

\input{sections_original/4_1_on_chain}
\input{sections_original/4_2_off_chain}
\input{sections_original/4_3_results}

\input{sections_original/5_1_data_gap}
\input{sections_original/6_conclusions}

%
%
%


\bibliographystyle{apalike2}
\bibliography{bibliography/references}

\newpage

\appendix

\input{sections_original/supplemental_material}

\end{document}

%% file: sections_original/0_abstract.tex

\begin{abstract}

%
Entities like centralized cryptocurrency exchanges fall under the business category of virtual asset service providers (VASPs). 
As any other enterprise, they can become insolvent. VASPs enable the exchange, custody, and transfer of cryptoassets organized in 
wallets across 
distributed ledger technologies (DLTs). Despite the public availability of DLT 
transactions, 
the cryptoasset holdings of VASPs
are not yet subject to systematic auditing procedures. In this paper, we propose an approach to assess the solvency of a VASP by cross-referencing data from three distinct sources: cryptoasset wallets, balance sheets from the commercial register, and data from supervisory entities. We investigate 24 VASPs registered with the Financial Market Authority in Austria and provide regulatory data insights such as who are the customers and where do they come from. Their yearly incoming and outgoing transaction volume amount to 2 billion EUR for around 1.8 million users. We describe what financial services they provide and find that they are most similar to traditional intermediaries such as brokers, money exchanges, and funds, rather than banks. Next, we empirically measure DLT transaction flows of four VASPs and compare their cryptoasset holdings to balance sheet entries. Data are consistent for two VASPs only. This enables us to identify gaps in the data collection and propose strategies to address them. We remark that any entity in charge of auditing requires proof that a VASP actually controls the funds associated with its on-chain wallets. It is also important to report fiat and cryptoasset and liability positions broken down by asset types at a reasonable frequency.

\vspace{16pt}
\textbf{JEL Classification:} C81, F31, G15, G20, G33, M41, 033

\textbf{Keywords:} \textit{Blockchain, Proof of Solvency, Virtual Asset, Cryptoasset, VASP, Accounting, Auditing, Regulation}

\end{abstract}

%% file: sections_original/1_introduction.tex

\section{Introduction}
\label{sec:intro}

In 2022, the cryptoasset sector experienced a crash driven by two major incidents that exposed the repercussions of inadequate regulation and accountability in the industry. In May, Terra's algorithmic stablecoin protocol experienced a stablecoin run, similar to a bank run, on its associated cryptoassets LUNA and UST \citep{Klages2021In,briola2022anatomy}. This triggered the bankruptcy of the crypto lenders Celsius and Voyager, and the hedge fund Three Arrows Capital \citep{econ2022ftx}. In November, the crypto trading platform FTX filed for bankruptcy, leading to BlockFi's downfall and bankruptcy consideration for Aax and Genesis\footnote{See \url{https://nyti.ms/3WUnEP7}, \url{https://bit.ly/3kIjegD}, and \url{https://on.ft.com/3XIogs8}.}. 
Even more recently, in June 2023, the U.S. Security and Exchange Commission (SEC) brought forward charges against some of the largest U.S.-based VASPs~\citep{sec2023coinbase,sec2023binance}.

These companies, and other centralized cryptoasset exchanges (CEXs) like FTX, fall under the broader definition of virtual asset service providers (VASPs). They facilitate financial activity involving virtual assets (VAs), such as their exchange for other VAs or fiat currencies, their custody and transfer via cryptoasset wallets, and portfolio management services for their customers \citep{fma2021aml,eu2018aml,mica2022proposal,FATF2021guidance}. 

\begin{figure}[t]
	\centering
	\includegraphics[width=0.9\textwidth]{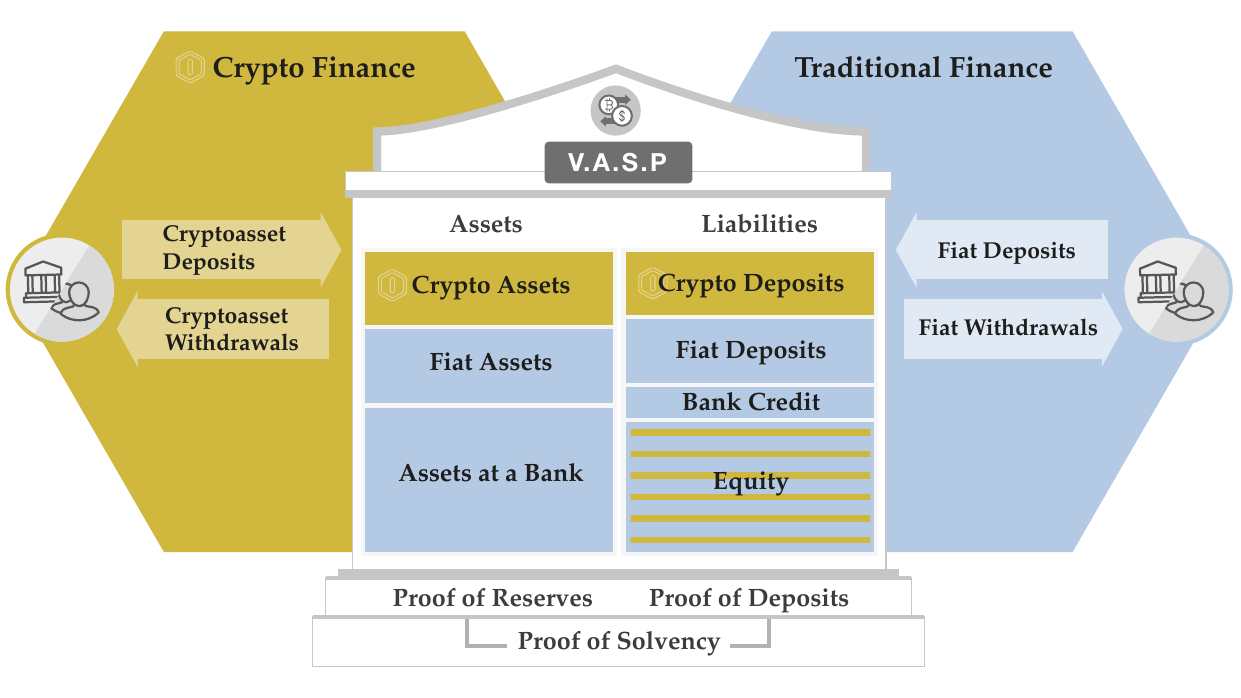}
	\caption{\textit{\textbf{Virtual Asset Service Provider.} VASPs hold virtual assets in custody, transfer them, and facilitate their purchase and sale against fiat currencies and other virtual assets. Customers can interact with them by depositing or withdrawing cryptoassets through DLT-based transactions, or fiat currency via commercial banks. }}
	\label{fig1:schematic_fig}
\end{figure}

%
As Figure~\ref{fig1:schematic_fig} shows, VASPs lie at the interface of the traditional and the crypto financial ecosystems, respectively called  \textit{off-chain} and \textit{on-chain} financial activity in jargon.
The aforementioned and other~\citep{moore2018toit} incidents that affected cryptoasset exchanges highlight a critical aspect of VASPs, i.e., the lack of proper accounting and business continuity concepts~\citep{zetzscheremaining}. While their \textit{off-chain} activities are audited according to generally accepted accounting principles, \textit{on-chain} assets are held in pseudo-anonymous cryptoasset wallets across multiple, possibly privacy-preserving DLTs~\citep{elbahrawy2017evolutionary} and are not yet systematically audited. 

Furthermore, whilst VASPs share several characteristics with traditional financial intermediaries, they are less regulated and their activities often lack transparency. Whether and how regulating them is an ongoing, highly controversial debate resembling a tug-of-war game; some argue for more VASP regulations~\citep{moser2019effective}, while others claim it would come at a substantial social cost and could be misunderstood as undeserved legitimacy\footnote{See e.g. \url{https://on.ft.com/3DgFHYD} and \url{https://bit.ly/3Y0JqBo}. }. A clear understanding of the financial functions VASPs provide, how they operate, and what risks are involved may provide guiding principles for regulators and policymakers. 

This paper proposes an approach for determining a virtual asset service provider's solvency status by measuring their cryptoasset holdings. By solvency, we mean that the total amount of assets held in custody is larger than the total amount of liabilities, whereby the difference is equity. We investigate the VASPs registered with the Financial Market Authority (FMA) in Austria in the context of the Anti-Money Laundering Act. 
We compare data from three distinct sources: we rely on publicly available DLT transaction records from the Bitcoin and Ethereum DLTs and use established algorithms \citep{androulaki2013evaluating,ron2013quantitative,Meiklejohn2016} to identify and cluster cryptoasset wallets likely controlled by the same entity. Then we reconstruct the VASPs' cryptoasset flows, compare their net positions to balance sheet data from the commercial register\footnote{The commercial register is a central, public directory. It contains important information about numerous companies. The primary purpose of the commercial register is to provide business transactions with the opportunity to obtain relevant information about the registered company.}, and complement them with supervisory data from FMA.

To the best of our knowledge, our work is the first that combines these distinct sources in a unified framework. Also, to our understanding, a consolidated approach to measuring the types of cryptoassets held by VASPs against their liabilities to customers still does not exist, although their activity is based on DLTs whose transactions are publicly auditable by design.

Moreover, we position VASPs in the landscape of financial intermediaries, by systematically comparing the services they offer to those of traditional financial service providers. 
While previous research has compared VASPs to banks~\citep{anderson2019bitcoin,dagher2015provisions}, we discuss why this comparison can be misleading. Our work provides the following contributions:

\begin{itemize}
    \item We study 24 Austrian VASPs and systematize the services they offer. We find that they are most similar to \textit{brokers}, \textit{money exchanges}, and \textit{funds}, rather than to \textit{banks};
    
    \item We provide regulatory data insights showing that their yearly incoming and outgoing transaction volume in 2022 amounted to 2 billion EUR for around 1.8 million users;
    
    \item We measure on-chain transaction flows for four VASPs and compare their holdings to balance sheet data from the commercial register. Data are consistent for two VASPs only; 
    
    \item We identify gaps in data collection practices and propose strategies to fill them: any entity in charge of auditing requires proof that a VASP actually controls the funds associated with its on-chain wallets; it is also important to report fiat and crypto asset and liability positions broken down by asset types, and at a reasonable frequency.
\end{itemize} 

Currently, supervisory auditing of VASPs does not fully exploit the public availability of DLT transactions. We believe our work provides valuable insights toward a better and more systematic assessment of their solvency, and might help make the process more effective and less error-prone. By comparing the VASPs cryptoasset holdings to balance sheet data, we show that the major issues are related to the different management of cryptoasset wallets in different DLTs, the lack of wallet addresses attribution data for VASPs, and the absence of breakdowns by cryptoasset types in balance sheets.

The paper is structured as follows. In Section~\ref{sec:litrev}, we introduce key background concepts and review the literature. Then we analyze VASPs and their features in Section~\ref{sec:background}. Section~\ref{sec:data_methods} describes the data, our measurement approach, and reports our results. In Section~\ref{sec:discussion} we discuss how the data gap can be reduced, while in Section~\ref{sec:conclusions} we draw conclusions. Our study follows an open-source approach and can be replicated on any other entity, provided the data on their cryptoasset wallets.

%% file: sections_original/2_1_background.tex

\section{Background and related literature}
\label{sec:litrev}

\subsection{Definitions - What is a VASP?}
\label{sec:defs}

While the term VASP has become increasingly common, its precise meaning and the specific activities that fall under this term still need to be clarified. 
We begin by providing the definition of VASPs according to the \citet{fma2021aml}, which follows the $5{th}$ EU AML directive~\citep{eu2018aml} and the Financial Action Task Force guidelines~\citep{FATF2021guidance}. According to this definition, a \textbf{Virtual Asset}, implemented on a distributed ledger technology, is~\citep[p.~VII]{fma2021aml}

\begingroup
\addtolength\leftmargini{0.1in}
\begin{quote}
	\textit{[...] a digital representation of value that is not issued or guaranteed by a central bank or a public authority, is not necessarily attached to a legally established currency and does not possess a legal status of currency or money, but is accepted by natural or legal persons as a means of exchange and which can be transferred, stored and traded electronically.}
\end{quote}
\endgroup

We note that in the ``Market in Crypto Asset Regulation'' (MiCA), the term ``crypto asset'' is used instead of virtual asset.  In our context, the two terms can be considered equivalent\footnote{MiCA also refers to crypto asset service providers (CASPs), rather than to VASPs. For the purpose of this work, VASPs and CASPs can be considered as synonymous as well.}.

\textbf{Virtual Asset Service Providers} are any natural or legal person that, as a business, conducts activities or operations for or on behalf of another natural or legal person. They can offer ``\textit{[...] one or more services}''~\citep[p.~VIII]{fma2021aml}, which we summarize in Table~\ref{tab:va_services}.

\begin{table}[h]
	\centering
	\input{tables/services.tex}
	\caption{\textit{\textbf{Description of the services provided by VASPs in Austria.}}}
	\label{tab:va_services}
\end{table}

VASPs lie at the interface of the traditional and the crypto financial ecosystems. The former encompasses financial activity with fiat currencies, i.e., legal tender money, and fiat assets, i.e., assets denominated in fiat currencies (similarly to cryptoassets being assets denominated in a cryptocurrency). It can rely on commercial banks and other traditional financial intermediaries. The latter entails financial activity executed on Distributed Ledger Technologies (DLTs) like the Bitcoin and Ethereum blockchains, and with cryptoassets such as bitcoin, ether, and the stablecoins tether (USDT), USD coin (USDC), or DAI. 

On the off-chain side, VASPs and customers interacting with them have strong identities; that is, the former need to register with regulatory bodies and the latter undergo identification processes such as KYC and AML5 compliance. On the on-chain side, activities involve weak identities~\citep{moser2013inquiry,ford2019rationality}: transactions, enabled by cryptographic keys, occur among pseudonymous counterparts, and the same entity can control multiple addresses.

We also note that VASPs differ from decentralized finance (DeFi) actors. This term indicates an emerging financial ecosystem built on DLTs that is non-custodial and does not require a central organization to operate~\citep{saggese2023defi}. 
VASPs are instead centralized intermediaries that provide interfaces to exchange cryptoassets via conventional IT systems, and transactions are not necessarily recorded on DLTs but at times are rather stored in private ledgers~\citep{aramonte2021defi,Auer2022cryptoshadow}.

\subsection{Proof of Solvency}

A company is solvent if the total amount of assets held in custody is larger than the total amount of liabilities, whereby the difference is equity.
Substantial documentation exists regarding incidents and exchange closures of VASPs~\citep{moore2018toit}, including recent events such as FTX's bankruptcy filing.
To increase transparency and foster trust,  several VASPs have recently disclosed lists of cryptoasset wallet addresses as a \textit{proof of reserve}, i.e., proof that they hold a given amount of assets.
However, such an approach alone does not constitute a valid \textit{proof of solvency} because it does not guarantee that VASPs have the financial resources to meet their current and future obligations\footnote{\textit{proof of reserves} and \textit{proof of solvency} are terms adopted in jargon. More technically, the latter is the capital cushion to fulfill liabilities and obligations against customers, i.e., the capital requirements. Proof of reserves were collected by projects such as DefiLLama, which gathered several CEX wallet addresses: \url{https://bit.ly/3KpdnHT}.}. First, a \textit{proof of deposits}, i.e., a verification of the customers' deposit amount, is needed as well~\citep{buterin2022cexs}. Second, in addition to revealing the existence of an address, it is necessary to prove control over the corresponding private key. Third, even this might not be sufficient, as colluding actors could lend each other cryptoassets to conduct one-time proof of reserves.\footnote{See e.g., \url{https://bit.ly/3XXBlgP} and \url{https://bit.ly/3DjaJiq}.} 

Data from the commercial register contains both information on the asset and liability side of VASPs balance sheets, and the fiat assets are audited according to generally accepted accounting principles. Therefore, in our context, it is sufficient to verify that the asset side is consistent with the cryptoasset holdings of a VASP to prove its solvency.

%% file: tables/services.tex
\renewcommand\tabularxcolumn[1]{m{#1}}
\begin{tabularx}{\linewidth}[h]{p{2.7cm}X}
	\textbf{Service} & \textbf{Description} \\
	\toprule
	\addlinespace[7pt]
	\textbf{Custodian} & Services to safeguard private cryptographic keys, to hold, store and transfer virtual assets on behalf of a customer (custodian wallet providers) \\
	\specialrule{0.001pt}{5pt}{7pt}
	\textbf{\makecell[l]{V2F-Exchange
	}} & Exchanging of virtual assets into fiat currencies and vice versa \\
	\specialrule{0.001pt}{-5pt}{7pt}
	\textbf{\makecell[l]{V2V-Exchange
	}} & Exchanging of one or more virtual assets between one another \\
	\specialrule{0.001pt}{-5pt}{7pt}
	\textbf{Payment} & Transferring of virtual assets\\
	\specialrule{0.001pt}{5pt}{7pt}
	\textbf{Issuance} & Provision of financial services for the issuance and selling of virtual assets\\
	\addlinespace[5pt]
	\bottomrule
\end{tabularx}

%% file: sections_original/2_2_trad_fin.tex

\subsection{Traditional financial intermediaries}
\label{sec:intermed}

VASPs allow customers to deposit and exchange assets, and can provide consulting and portfolio management services for their customers, often holding funds on behalf of their customers \citep{anderson2019bitcoin}. Therefore, they share several characteristics with traditional financial intermediaries.
Here we describe the ones that are most important in our context and their main economic functions. A comprehensive description of traditional financial intermediaries can be found in \citet{howells2008economics} and \citet{cecchetti2014money}.

The two primary financial intermediary categories are deposit-taking institutions (DTIs) like banks and non-deposit-taking institutions (NDTIs) such as brokerage firms, mutual funds, and hedge funds. Other DTIs include building societies (UK), savings and loan societies (US), and mutual and cooperative banks (DE, FR). The main difference is that DTIs issue loans and use their liabilities as official money~\citep{howells2008economics}. Consequently, an increase in DTI business ultimately increases the money supply in an economy.

\textit{Banks}, by far the most important DTIs, pool small savings to make large loans. They also provide liquid deposit accounts, access to the payment system, and screen and monitor borrowers.
Among NDTIs, \textit{brokerage firms} facilitate access to trading in financial instruments. They offer custody and accounting services for customer investments and are additionally involved in the clearing and settlement of trades. \textit{Mutual funds}, including exchange-traded funds (ETFs), sell shares to customers and invest in a diverse range of assets offering access to large, diversified portfolios. \textit{Hedge funds} operate as financial partnerships, often requiring accredited or high-net-worth investors. They pool savings to earn returns through actively managed investment strategies, including derivatives, arbitrage, and short sales.

%% file: sections_original/2_3_literature.tex
\subsection{Literature}

The academic literature on VASPs is vast and primarily focuses on cryptoasset exchanges, highlighting the central role they have in the crypto ecosystem~\citep{makarov2021blockchain,lischke2016analyzing}.  
Recent studies show that most of the trading on cryptoasset markets happens off-chain on CEXs~\citep{auer2022banking,brauneis2019high}; according to \cite{makarov2021blockchain}, 75\% of the bitcoin transactions involve exchanges or exchange-like service providers.
CEXs play a major role also as they facilitate price discovery~\citep{brandvold2015price}.
Scholars exploited price time series from the largest exchanges to investigate the price formation dynamics~\citep{kristoufek2015main,katsiampa2017volatility,li2017technology} and estimate the fundamental value of cryptoassets~\citep{cheah2015speculative,kristoufek2019bitcoin}.
Other studies used instead exchange-based data to investigate topics such as market (in)efficiency~\citep{urquhart2016inefficiency,kristoufek2018bitcoin}, the behavior of bitcoin as a
currency or asset~\citep{glaser2014bitcoin,yermack2015bitcoin}, the effects of cross-listing on returns~\citep{benedetti2021returns}, as well as arbitrage, both across exchanges~\citep{makarov2020trading} and within one exchange alone~\citep{saggese2023arbitrageurs}. Exchange data were also used to study market microstructure aspects, such as price jumps~\citep{scaillet2020high}, or market liquidity~\citep{brauneis2022bitcoin}.
Another relevant strand of literature investigates risks associated with CEXs such as price manipulation~\citep{gandal2018price}, susceptibility to attacks~\citep{feder2017impact}, wash trading~\citep{chen2022cryptocurrency} and data fabrication~\citep{cong2022crypto}.

Previous studies have provided taxonomies or categorizations of crypto financial intermediaries~\citep{kazan2015value,blandin20203rd} and compared them to traditional ones~\citep{aramonte2021defi}. In \citet{fang2022cryptocurrency}, the authors define cryptoasset trading and survey the related works. Our work differs in that we base our categorization of virtual asset service providers on the legal definition of the Austrian Financial Market Authority. Then, we identify the financial functions that VASPs offer and provide an overview of the different types of VASPs. 

Close to our work, \citet{decker2015making} and \cite{dagher2015provisions} implemented a software-based solution to automate the audit of centralized Bitcoin cryptoasset exchanges.
Other works focused instead on proof of reserves for less relevant DLTs~\citep{dutta2019mprove,dutta2019revelio,dutta2021mproveplus}.
Our work differs as it is based on an empirical approach that cross-references multiple different sources of information (cryptoasset wallets, balance sheet data from the commercial register, and information from supervisory entities), and because it focuses on the two most relevant blockchains (Bitcoin and Ethereum), but can be extended to others. 


%% file: sections_original/3_1_VASPs.tex

\section{VASPs: A Closer Examination}
\label{sec:background}

Using information from the Austrian Financial Market Authority (FMA), we first describe what financial services they offer and what cryptoassets they support. Next, we complement the FMA data with additional public information collected from the VASPs websites to group them based on similarity scores. Finally, we compare their economic functions highlighting similarities and differences to traditional financial intermediaries.

\subsection{The Austrian VASP landscape}\label{sec:vasps}

\begin{figure}
\centering
\begin{subfigure}{.45\textwidth}
  \centering
  \includegraphics[width=0.9\linewidth]{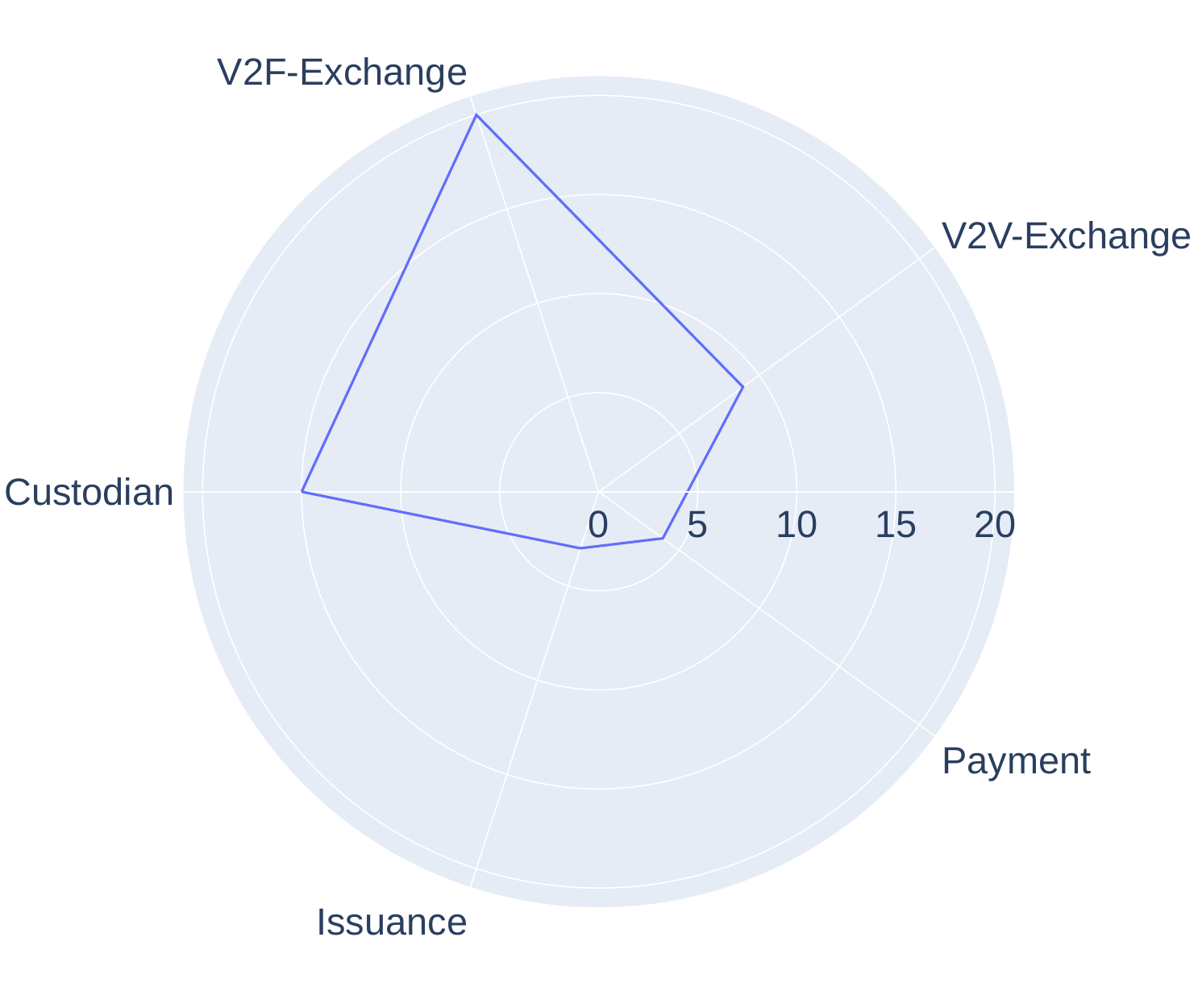}
  \caption{}
  \label{fig:VASP_FMA_features}
\end{subfigure}%
\begin{subfigure}{.45\textwidth}
  \centering
  \includegraphics[width=0.9\linewidth]{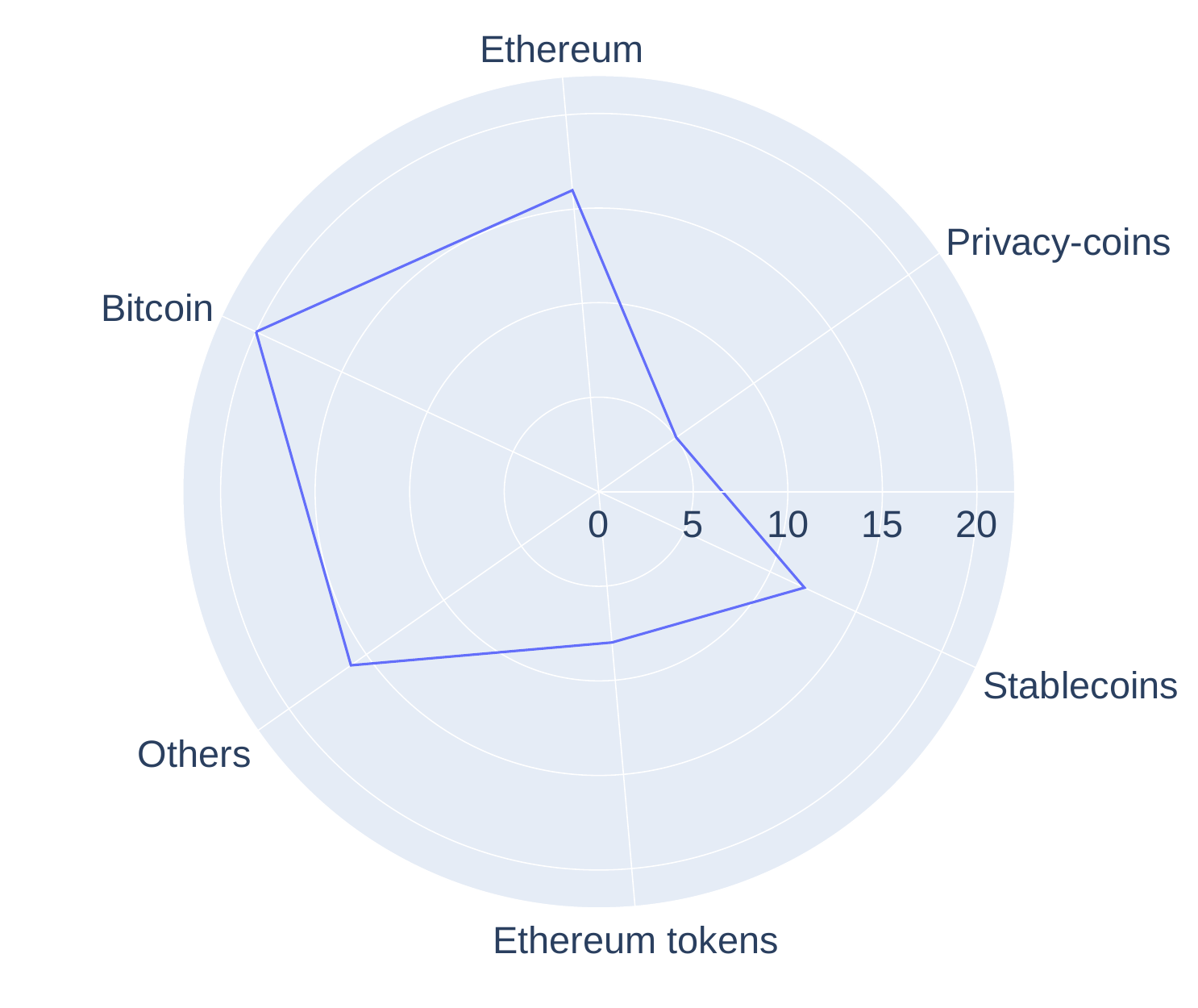}
  \caption{}
  \label{fig:VASP_tokens}
\end{subfigure}
\caption{\textit{\textbf{The Austrian VASP landscape}. Subfigure~(\subref{fig:VASP_FMA_features}) shows the number of VASPs registered for each of the five service categories described above. Most of the VASPs offer V2F-Exchange (N = 20), and offer more than one service, such as custody (N = 15). Subfigure~(\subref{fig:VASP_tokens}) reports how many VASPs offer services related to bitcoin (N = 20), ether (N = 16), and other relevant cryptoassets. Most VASPs exploit multiple DLTs.}}
\label{fig:VASPs}
\end{figure}

VASPs in Austria are supervised by the Financial Market Authority (FMA) under the Anti-Money Laundering Act. In December 2022, 24 VASPs were registered in the FMA database\footnote{\url{https://bit.ly/3kMUkwg}}.  

Figure~\ref{fig:VASP_FMA_features} shows the aggregate number of VASPs registered for each service described in Table~\ref{tab:va_services}. The vast majority of them ($N = 20$) offers \textit{V2F-Exchange}, i.e., services to exchange virtual assets and fiat currencies; nine also facilitate the exchange from and to other virtual assets (\textit{V2V-Exchange}). In most cases, customer funds are or can be kept in custody by the VASP ($N = 15$). Finally, only a few of them are legally authorized to transfer virtual assets and to issue and sell them (respectively services \textit{Payment}, $N = 4$, and  \textit{Issuance}, $N = 3$).
Additional details on the number of services offered per VASP are reported in Appendix A.

Figure~\ref{fig:VASP_tokens} shows what virtual assets are used by the Austrian VASPs. We follow the taxonomy described in~\citet{saggese2023defi} to aggregate cryptoassets into five categories. We could retrieve reliable information for 20 VASPs out of the 24 in the FMA database. Notably, all VASPs offer services related to bitcoins ($N = 20$). More than 75\% support Ethereum ($N = 16$), and the latter typically also support Ethereum tokens, i.e., ERC-20 and (or) ERC-721 compatible non-native tokens, and stablecoins (respectively $N = 8$ and $N = 12$). A limited number of VASPs also provide services related to privacy-focused cryptoassets (i.e., Monero, Dash, Zcash). Finally, several VASPs also support tokens native to other DLTs (e.g., Litecoin or Cardano).

\begin{figure}
	[t]
	\centering
	\includegraphics[width=0.9\textwidth]{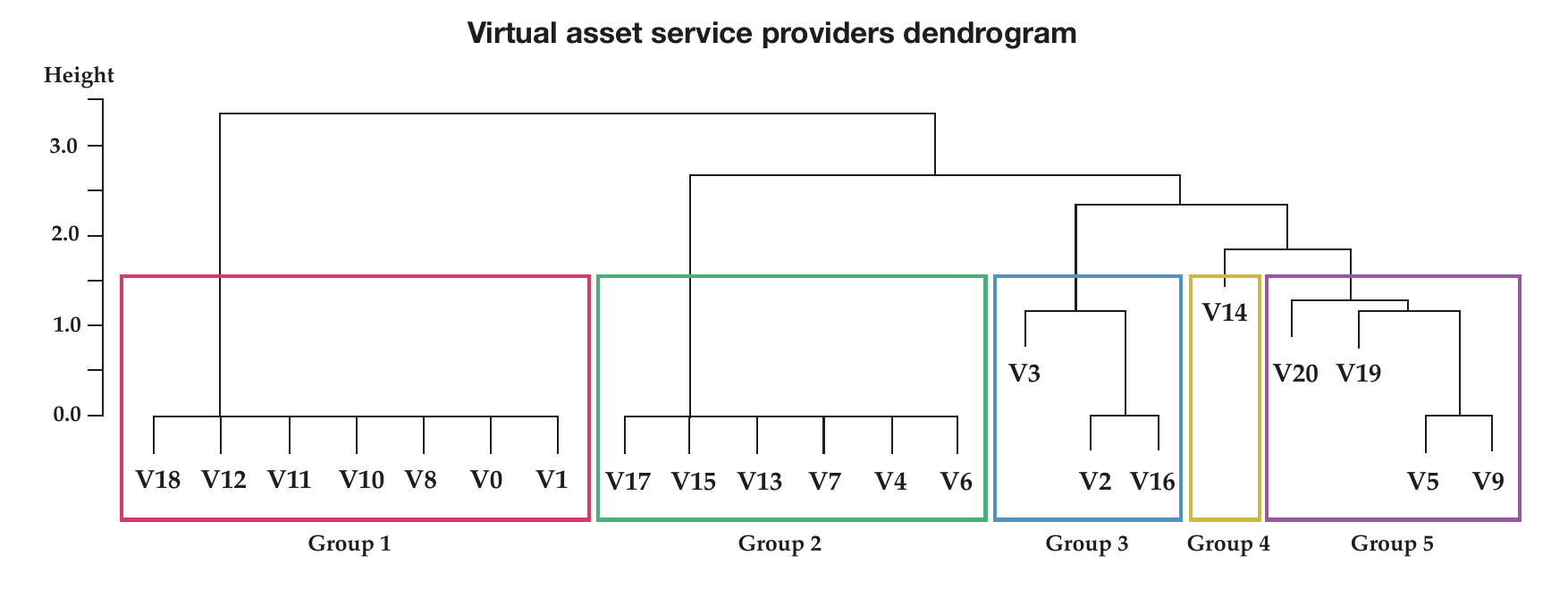}
	\caption{\textit{\textbf{VASPs categorization by their service offering.} We use a hierarchical agglomerative clustering approach to categorize VASPs. The two largest groups are VASPs that facilitate the exchange of virtual assets without offering custody ( \protect\tikz \protect\draw[black!20!red, line width=1pt] (0,0) rectangle (0.3,0.2);
	) and offer consulting services ( \protect\tikz \protect\draw[black!40!green, line width=1pt] (0,0) rectangle (0.3,0.2);
	). The others offer custody and exchange services ( \protect\tikz \protect\draw[white!30!violet, line width=1pt] (0,0) rectangle (0.3,0.2);
	), are payment processors (
	\protect\tikz \protect\draw[black!30!yellow, line width=1pt] (0,0) rectangle (0.3,0.2);
	), or implement trading platforms (
	\protect\tikz \protect\draw[black!30!cyan, line width=1pt] (0,0) rectangle (0.3,0.2);
	).}}
	\label{fig:dendrogram}
\end{figure}

In addition to FMA data, we collect additional public information documented on their websites. Our aim is to categorize VASPs by their service offering. We construct categorical variables that indicate whether the VASP offers custody services, facilitates payments, allows users to exchange cryptoassets, implements a trading platform, or offers consulting or investment services. We consider 21 VASPs for which we could gather sufficient information. Data for each (anonymized) VASP are reported in~\ref{sec:supplemental}.
Whilst the sample is small and the features are few, to ensure consistency and objectivity in categorizing VASPs we exploit an unsupervised learning method.
We aggregate them using the hierarchical agglomerative clustering (HAC) method~\citep{murtagh2012algorithms}. With this bottom-up approach, objects are iteratively clustered based on their similarity. The two main parameters of HAC are the distance among objects, and the linkage method, i.e., the distance used to merge groups. In our setting, we select the Euclidean distance and the Ward metric, and distances are iteratively computed using the Lance–Williams update formula. Results are similar when using other parameters. We report our classification in Figure~\ref{fig:dendrogram}. It categorizes VASPs into five clusters: the first one (red rectangle, $N = 7$) includes VASPs that do not keep customers' funds in custody, and only facilitate the exchange of virtual assets for fiat currencies and (or) other virtual assets. Some of them automate the process through physical vending machines. Thus the first left branch mainly separates VASPs that offer custody and those that do not. We identify them as ``Group~1''.

The green rectangle identifies VASPs providing investment advice and/or portfolio management in addition to custody services ($N = 6$). They propose investing strategies, give advice on portfolio management and coin selection services, and in some cases, lend customer funds. These VASPs are referred to as ``Group~2''.

The purple rectangle (``Group~5'') aggregates VASPs that act as cryptoasset custodians. Typically, they also facilitate the exchange of cryptoassets and are similar to VASPs in the blue rectangle (``Group~3''). These VASPs in addition provide customers with an internal trading platform, manage and match orders in a private limit order book, and update their account balances in cryptoasset or fiat money when trades are executed. Trades executed in private ledgers do not affect the public distributed ledgers unless the customers withdraw cryptoassets from the service. Such VASPs play an essential role in the crypto-financial system. As a result of the matching mechanism for demand and supply, these are the platforms where price formation takes place. The other VASPs derive their offered prices from other platforms as an exogenous variable. In the following, we consider these two as a single group (i.e., Group~3).

All the VASPs in the groups described above are cryptoasset centralized exchanges, or CEXs. The remaining VASP in the yellow rectangle is instead a payment processor service. It offers solutions to facilitate the purchase and sale of commodity goods with cryptoassets; such VASPs play a minor role in the crypto ecosystem.

%% file: sections_original/3_2_comparison.tex

\subsection{A comparison with traditional financial intermediaries}

\begin{figure}[t]
	\centering
	\includegraphics[width=\textwidth]{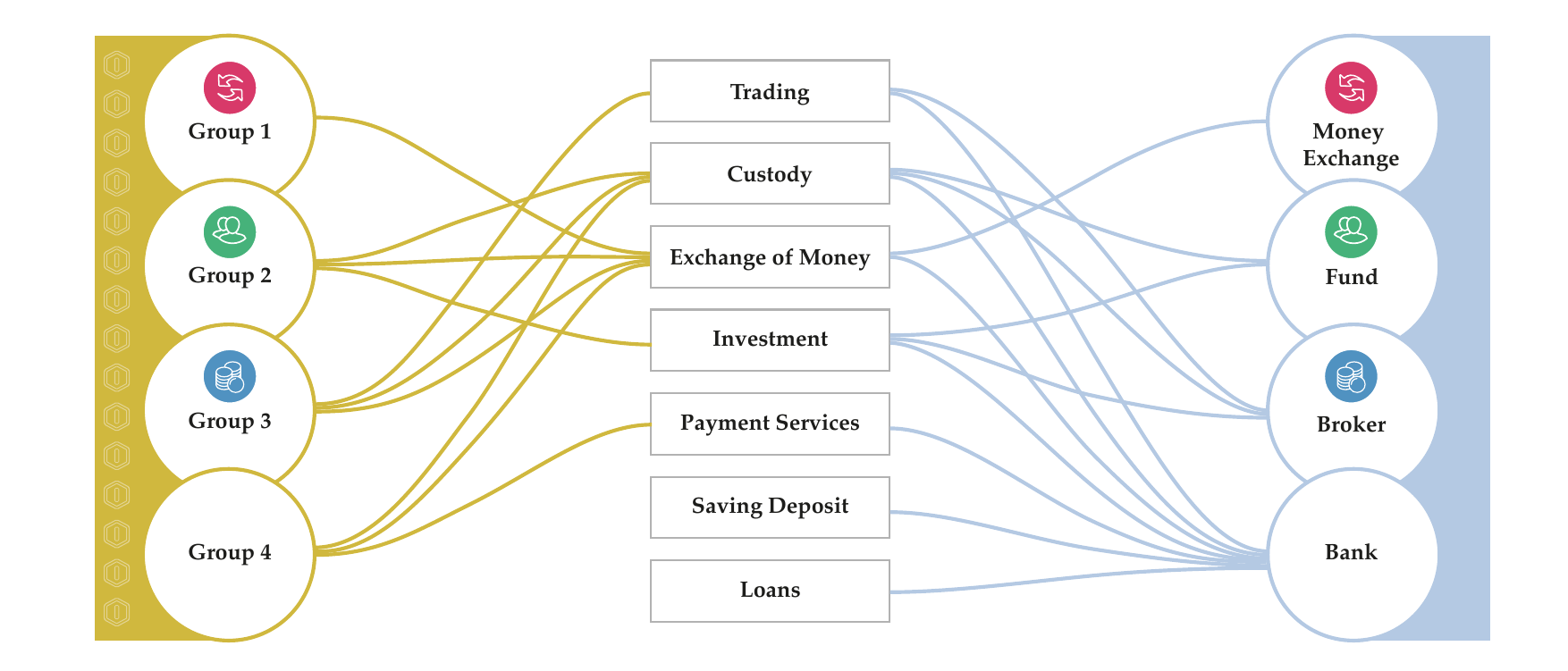}
	\caption{\textit{\textbf{Comparison of traditional financial intermediaries with VASPs.} Circles on the left represent VASPs, divided into groups as described in Figure~\ref{fig:dendrogram}, while on the right are traditional financial intermediaries. Links point to the financial functions offered by each financial intermediary. VASPs are most similar to money exchanges, brokers, and funds, rather than banks. The colors in the circles highlight what traditional intermediary each group is most similar to.}} 
	\label{fig:comparison}
\end{figure}

Having outlined the landscape of VASPs in Austria, we are now interested in understanding how they differ from traditional financial intermediaries. 
Figure~\ref{fig:comparison} stylizes the traditional financial intermediaries on the right and the VASPs on the left. In the middle, rectangles represent the primary economic services, and links indicate what services each intermediary category offers. The comparison shows that an analogy with traditional intermediaries exists for three out of the four groups described in Figure~\ref{fig:dendrogram}. More specifically, VASPs in group 1 operate similarly to \textit{money exchanges}. Indeed, the only service they offer is to buy and sell virtual assets for customers. VASPs in group 2 provide investment services to their users, akin to \textit{funds}. Third, groups 3 (and 5) include VASPs allowing users to trade, keep their funds in custody, and thus act as \textit{brokers}, connecting buyers and sellers to facilitate a transaction. The last group that provides payment services can be compared to payment processor systems.

Interestingly, we find that the comparison of VASPs to banks can be misleading: while the two share overall several financial services, such as exchanging money, trading, or investing, banks also enable customers to open loan positions with the funds they hold and to open savings and deposit positions. On this note, we mention that some VASPs have recently acquired an e-money institution license. However, that does not qualify them to offer bank-type financial services automatically. First, e-money institutions do not have the same supervisory requirements as traditional banks. Second, they do not necessarily offer bank-type financial services --- they can e.g. use the license only to process their fiat payments. Further information on the taking up, pursuit, and prudential supervision of the business of electronic money institutions can be found in the Directive 2009/110 of the European Commission \citep{eu2009directive}.

%% file: sections_original/4_1_on_chain.tex

\section{Measuring VASPs Cryptoasset Holdings}
\label{sec:data_methods}

After describing the VASPs service offerings, we now move on and devise an approach to empirically assess their solvency by correlating data from multiple on-chain and off-chain sources.
The underlying intuition is that, by quantifying the cryptoassets held on-chain by one VASP, we should be able to verify the numbers reported in the balance sheets. Furthermore, it is sufficient to measure the asset side, because on the liability side cryptoassets are either customer liabilities or equity. Since balance sheet assets minus liabilities are equal to equity, our approach serves as a first proof of solvency.


We first discuss which DLTs we analyze, motivate our choice, and document our approach to reconstruct the VASPs net positions by extracting the data from the two most relevant DLTs, Bitcoin and Ethereum. VASPs wallet addresses are extracted from a large collection of public attribution tags, or identified by executing manual transactions, and have not been revealed by the VASPs themselves. Next, we describe the balance sheet data from the commercial register. We concentrate our empirical analysis on four VASPs whose wallets appear in the attribution tag collection and that have published their balance sheets consistently over time, allowing to compare on-chain cryptoasset holdings to balance sheets. Their market share is around 99\% of the total market share. 

\subsection{On-chain data}
\label{sec:data}

DLTs can be divided into two major typologies based on their conceptual design. They either follow the Bitcoin-like Unspent Transaction Output (UTXO) or the Ethereum-like account model. Both support by design a native token, like bitcoin or ether. The latter, by enabling the deployment of arbitrary smart contracts, also supports issuing non-native tokens such as the stablecoins USDT, USDC, and DAI.

We begin by gathering the transaction history of the two most relevant DLTs, Bitcoin and Ethereum, from their origin to the 3$^{rd}$ of April 2022\footnote{The time frame can be extended to 2022 to include the balance sheet of upcoming years when available.}. We focus on the Bitcoin and Ethereum ledgers for the following reasons: first, as shown in Section~\ref{sec:background}, all VASPs operate with bitcoins and in most cases also with ether. Cryptoassets deployed on other DLTs are less relevant. Second, bitcoin, ether, and the stablecoins USDT and USDC alone account for more than 70\% of the total cryptoasset market capitalization, and these are also the most traded and held cryptoassets by CEXs customers\footnote{See \url{https://coinmarketcap.com/charts/} and \url{https://coinmarketcap.com/rankings/exchanges/}}. Third, while stablecoins like USDT are deployed on multiple smart contract-compatible ledgers\footnote{see, e.g., USDT \url{https://bit.ly/3YSYNwR} and USDC \url{https://www.circle.com/en/multichain-usdc}} and currently deploy significant amounts of tokens also in other DLTs\footnote{\url{https://tether.to/en/transparency/}.}, Ethereum is historically the most relevant one.

We implement two approaches to extract on-chain VASP-related information for the UTXO-based and the account-based DLTs. The entities that operate on the Bitcoin blockchain interact with each other as a set of pseudo-anonymous addresses. We exploit known address clustering heuristics~\citep{androulaki2013evaluating,ron2013quantitative,Meiklejohn2016} to associate addresses controlled by the same entity\footnote{New addresses can be created in each transaction. However, if they are re-used across transactions, they can be linked and identified as belonging to the same entity.}. Furthermore, we exploit a collection of public tagpacks, i.e., attribution tags that associate addresses with real-world actors, to filter the clusters associated with any of the VASPs considered in our study. We expanded the dataset by conducting manual transactions with the VASPs in our sample (further details are discussed in~\ref{sec:supplemental}, where we also report a list of the addresses used). We identified 88 addresses and their corresponding clusters associated with four different VASPs.

To reconstruct their net positions, we filter the Bitcoin transaction history and select only the transactions in which the sender or recipient is an address associated with the four VASPs. In total, we consider \num{1574125} Bitcoin transactions.

We use a different approach for the Ethereum DLT. An Ethereum address identifies an account whose state is updated via state transitions through transactions. The account state stores information about the balance and the number of transactions executed, maintaining thus a historical database. While approaches for address clustering have been devised for Ethereum as well~\citep{victor2020address}, in practice, addresses are typically reused. We thus extract all relevant information by running a full Erigon Ethereum archive node~\citep{eirgon2022node}. Similarly to the previous approach, we exploit attribution tags and manual transactions to identify the addresses associated with VASPs. In total, we identified nine relevant addresses associated with three different VASPs. We proceed by querying the state of each account, from the beginning of the Ethereum transaction history (block 0) to the 3$^{rd}$ of April 2022, every \num{10000} blocks. In addition to the ether balance, we collect data on the address balance for the tokens USDT, USDC, DAI, wETH, wBTC. The list of ground-truth addresses is reported in the appendix.

We remark that our attribution dataset contains more than \num{265000000} deanonymized Bitcoin addresses, covering more than 24\% of the total number of existing Bitcoin addresses. In addition, \num{278244} tagged Ethereum addresses cover 0.11\% of the existing addresses. The former identifies around \num{3000} entities active in the Bitcoin ecosystem, the latter more than \num{25000} Ethereum entities.

%% file: sections_original/4_2_off_chain.tex

\subsection{Off-chain data}

We collect balance-sheet data for 17 Austrian VASPs through the Austrian Commercial Register. We construct an unbalanced panel data\footnote{i.e., time observations are different for different VASPs.} starting from 2014 to 2021. Ultimately, in our empirical analysis, we use the data of four Austrian VASPs for which we can identify on-chain and off-chain data. Our variable of interest is a firm-level measure of crypto asset holdings. Some firms describe their crypto asset holdings as explicit balance-sheet items; for other firms that aggregate them with other items we construct a variable that approximates the corresponding crypto asset holdings from their described asset items. The balance sheet does not allow us to distinguish between cryptoasset holdings such as ether and bitcoin. The variable \emph{crypto asset holdings} in form of red markers in Figure~\ref{fig:VASP_A_all}, Figure~\ref{fig:VASP_B_eth}, Figure~\ref{fig:VASP_C_all} and Figure~\ref{fig:VASP_D_BTC} represents those balance-sheet items. 

%% file: sections_original/4_3_results.tex

\subsection{Comparing on- and off-chain data}

\begin{figure}[t]
	\centering
	\includegraphics[width=0.8\textwidth]{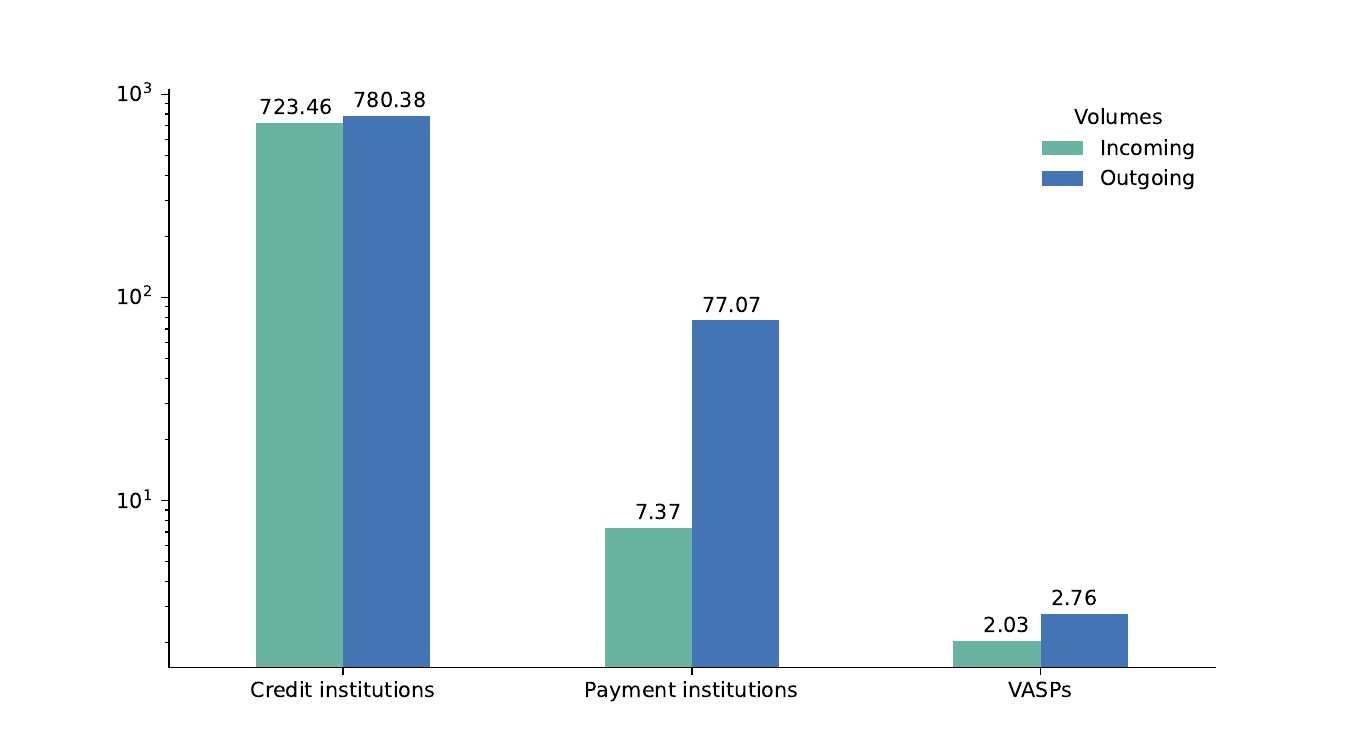}
	\caption{\textit{\textbf{Transaction volumes of Austrian VASPs and other financial intermediaries.} The incoming and outgoing transaction volumes of VASPs are respectively one order and two orders of magnitude smaller than those of payment institutions and credit institutions.}}
	\label{fig:trans_vol}
\end{figure}

Supervisory data from FMA show that in a 12-month period (roughly 2021 until 2022 due to varying reporting dates for VASPs), the transaction volume of virtual assets converted to EUR conducted by VASPs registered in Austria amounts to 2.03 billion incoming transaction volume and 2.76 billion outgoing. The transaction volume is computed as the sum of the transactions related to customer relationships only. As Figure~\ref{fig:trans_vol} shows, in comparison, during the same time we observed a transaction volume for credit institutions of 723.46 (incoming) and 780.38 (outgoing) billion and of 7.37 (incoming) and 77.07 (outgoing) billion for payment institutions.

Table~\ref{tab:customers} reports additional supervisory data from FMA on the number of VASP customers by residence and legal form. A VASP customer refers to a natural or legal person, who has opened an account and gone through a validated KYC process with the particular VASP. The rows distinguish natural persons, i.e., individuals, and legal persons, i.e., entities with legal rights. 
Customers are further divided by jurisdiction: the first column indicates the number of Austrian customers, while the second one reports the number of customers in the European Union, excluding Austrians (we note that customers are never counted in two columns). 
The subsequent columns identify customers by jurisdictions that are respectively offshore financial centers \citep{imf2019offshore}, subject to embargo \citep{wko2020embargo}, and under increased monitoring (grey list; \citealp{FATF2022grey}). The last columns respectively aggregate all remaining countries and report the total number of users. 
The assignment works in such a way that countries that appear in several lists will be assigned to the group that bears the greater risk.
Total customers are 1.79 million, and they are mainly natural persons. The vast majority are Austrian or members of the European Union (respectively around \num{327000} and N = \num{1279300}).
We note that this number might include customers who created an account but never transacted, i.e. the count is not weighted by transaction number. Furthermore, the same customers can have accounts at multiple VASPs. Customers from subsidiaries and inactive are excluded.

\begin{table}[t]
	\footnotesize
	\input{tables/customers.tex}

	\caption{\textit{\textbf{VASP customers residency in different jurisdictions.} We report figures for natural persons (top) and legal persons (bottom). Customers are never double counted; e.g., the first column reports the number of Austrian customers, while the second reports European Union members excluding Austrians. We further distinguish customers by jurisdictions that are offshore, subject to embargo, and under increased monitoring (``grey list''). The last columns aggregate all other jurisdictions (Other) and report the total number of customers (Total).} Source: supervisory data from FMA.}
	\label{tab:customers}	
\end{table}

The four entities we study cover around 99\% of the Austrian VASP transaction volumes measured in total assets. Consistently with the labels introduced in Figure~\ref{fig:comparison}, we denote them as \VASPA, \VASPD, \VASPC, and \VASPB and hide their real names to avoid that the corresponding VASPs can be directly recognized in our study. They are representative of different VASP groups (i.e., money exchanges, brokers, and brokers with trading platforms).

\subsubsection{\VASPA} 

\begin{figure}[h]
	\centering
	\includegraphics[width=\textwidth]{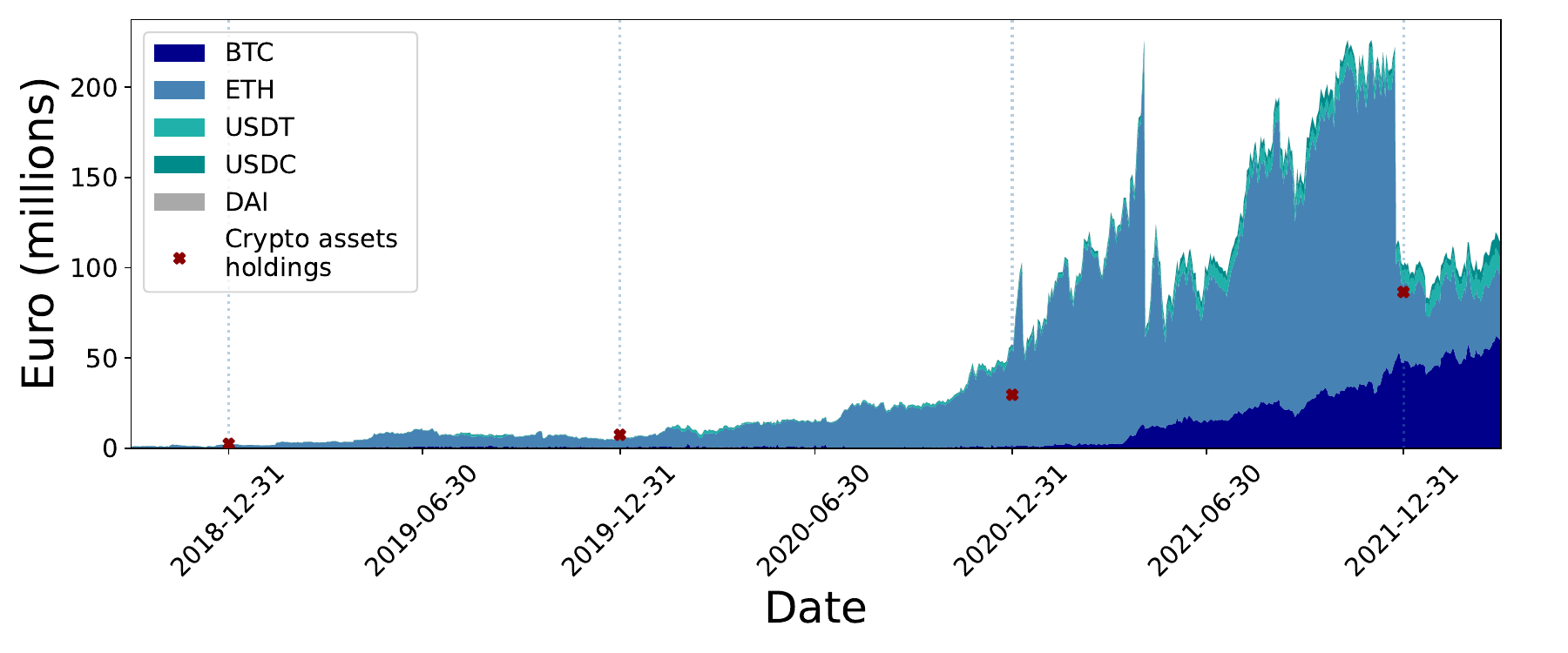}
	\caption{\textit{\textbf{Estimation of the cryptoasset holdings of \VASPA.} Colors correspond to different cryptoassets: bitcoin in dark blue, ether in light blue, USDC in dark green, USDT in light green, and DAI in gray. Red markers indicate the cryptoasset holdings declared in the balance sheet data at the end of each year for the period 2018 to 2021.}}
	\label{fig:VASP_A_all}
\end{figure}

\paragraph{Observations} We report the values for \VASPA in Figure~\ref{fig:VASP_A_all}. In this and the subsequent plots, the bitcoin holdings are in dark blue, ether in light blue, USDC in dark green, USDT in light green, and DAI in gray. The dots represent the cryptoasset holdings declared in the balance sheet data at the end of each year for the period 2018 to 2021. This VASP implements a trading platform and falls within group 3. 

The cryptoasset holdings identified on-chain correspond to 75.59\% of the cryptoassets declared in the balance sheet at the end of 2018, 66.68\% at the end of 2019, 194.56\% at the end of 2020, 116.79\% at the end of 2021. The amount of bitcoin increased significantly after April 2021, and the largest amount of tokens is held in ether.

\paragraph{Findings} Overall, the two sources of information point in the same direction. Interestingly, after 2020, the on-chain activity is higher than what the balance sheet reports. A possible interpretation is that the cryptoassets in excess represent equity or private funds. \VASPA reports well-separated balance sheet positions, allowing us to compute precisely the amount of cryptoasset holdings.

\subsubsection{\VASPB} 

\begin{figure}[h!]
	\centering
	\includegraphics[width=\textwidth]{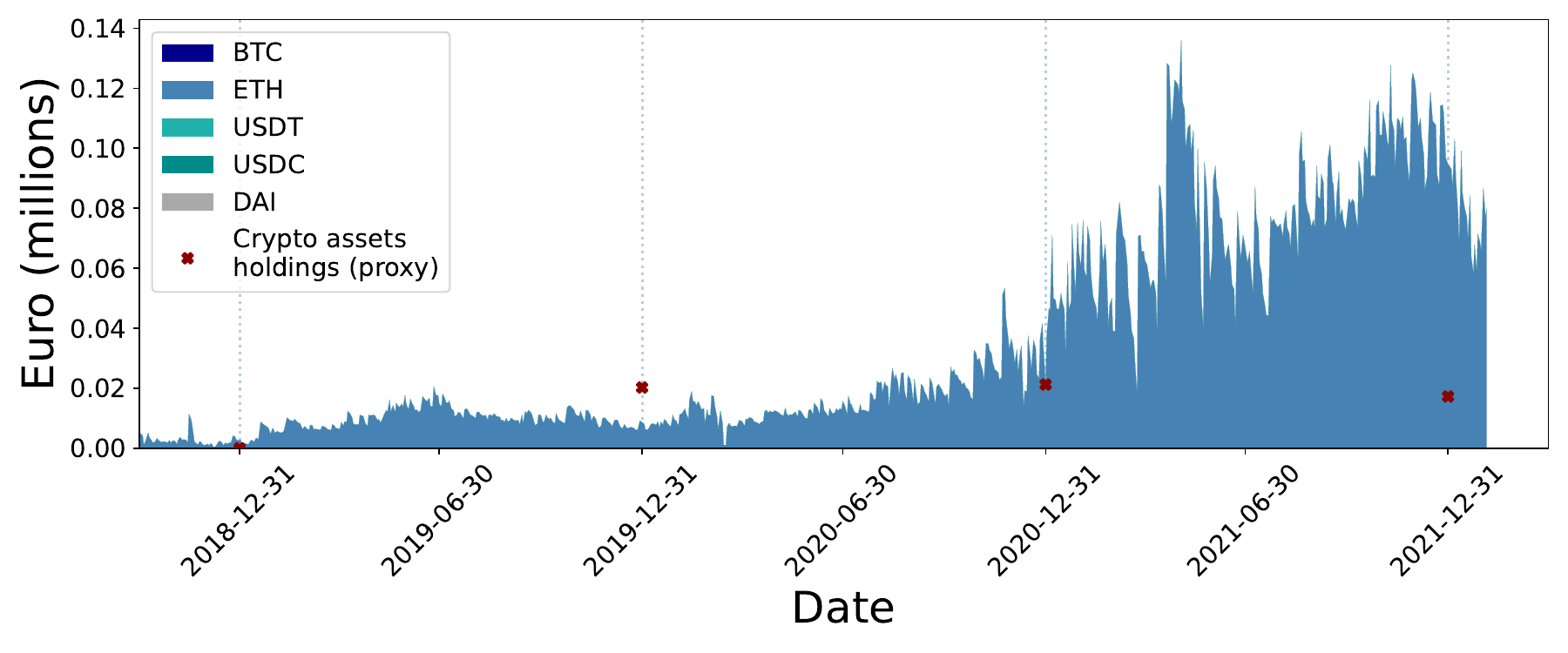}
	\caption{\textit{\textbf{Estimation of the cryptoasset holdings of the \VASPB.} On-chain and off-chain data correspond until the end of 2020.  All reported assets are ether. Balance sheet data are a proxy, as cryptoassets are aggregated with other items in the balance sheet.}}
	\label{fig:VASP_B_eth}
\end{figure}

\paragraph{Observations}
Figure~\ref{fig:VASP_B_eth} shows the cryptoasset holdings of \VASPB. It is a non-custodial VASP that provides exchange services based both on ether and bitcoin. The cryptoassets measured on-chain are partially comparable with those reported on the balance sheets (42.59\% at the end of 2019, 102.45\% at the end of 2020, but  549.38\% at the end of 2021).

\paragraph{Findings}

Similarly to \VASPA, on-chain activity is higher than the value reported on the balance sheet after 2020. As expected, the amount of cryptoasset holdings is small, as the VASP is non-custodial, and exceeds 100K EUR only after 2021. All reported assets are ether: the absence of stablecoins is expected, as this VASP trades bitcoin, ether, and a few other cryptoassets. However, we could not identify bitcoin flows from or to their wallets in the time frame we considered. To identify the addresses associated with this VASP, we relied on manual transactions: re-identification attacks are a possible strategy to collect attribution tags. While this strategy is effective for Ethereum accounts, the Bitcoin addresses we gathered identify the VASP activity dating back to November 2022 only, thus outside of the time frame we considered. 

Regarding balance sheet data, we note that the values, in this case, are a proxy: cryptoassets are aggregated with other items in the balance sheet.

\subsubsection{\VASPC} 

\begin{figure}[h!]
	\centering
	\includegraphics[width=\textwidth]{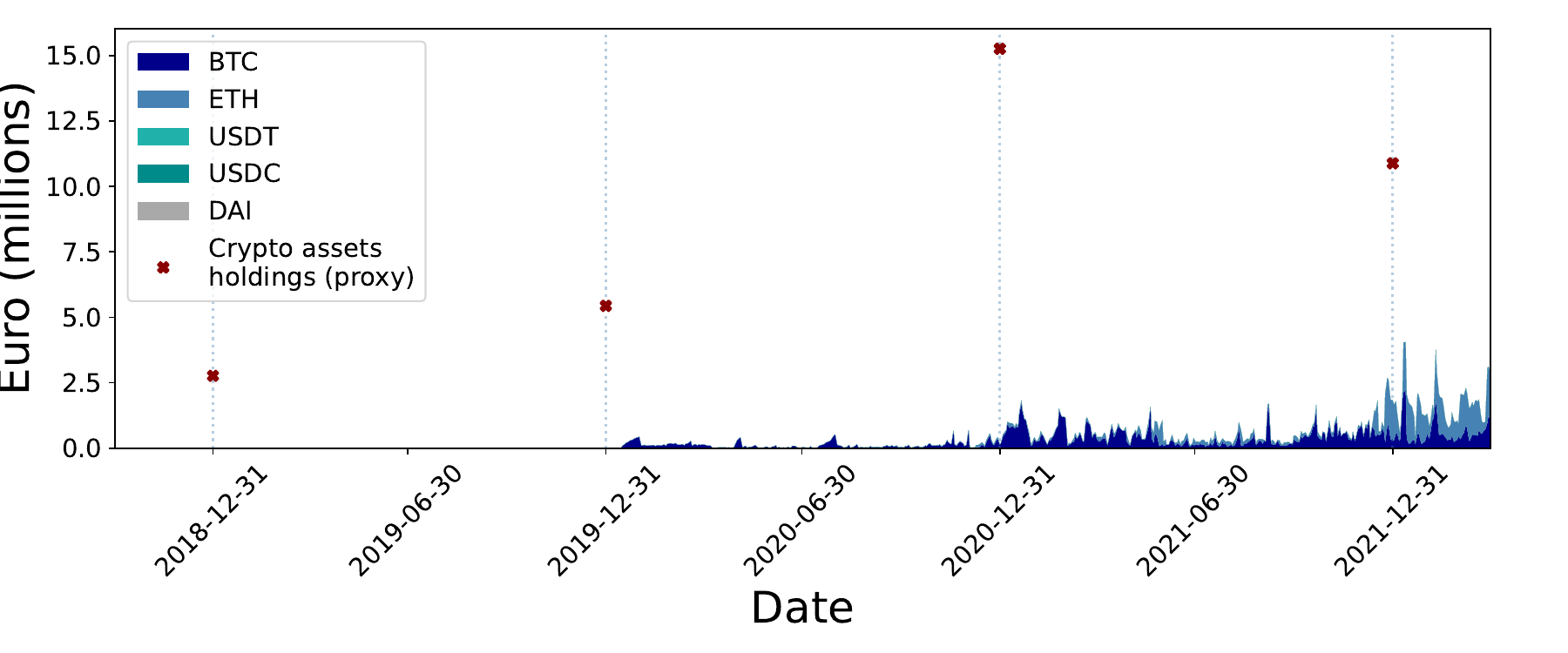}
	\caption{\textit{\textbf{Estimation of the cryptoasset holdings of \VASPC.} The cryptoasset holdings cover only a small fraction of the funds declared in the balance sheets.}}
	\label{fig:VASP_C_all}
\end{figure}

\paragraph{Observations}
\VASPC is shown in Figure~\ref{fig:VASP_C_all}. It is categorized in group 5 in Subsection~\ref{sec:vasps}. Unlike the previous cases, the cryptoasset holdings cover only a tiny fraction of the funds declared in the balance sheets; in the best case, i.e., at the end of 2021, we can identify on-chain only 16.85\% of the total cryptoassets reported in the balance sheet.

\paragraph{Findings}

A possible explanation for the discrepancy is that our dataset might include only \textit{hot wallets}, i.e., addresses used to conduct daily operations such as the deposit and withdrawal, but not the \textit{cold wallets}, i.e., addresses that control the large majority of customers funds and that are subject to stricter security measures. An alternative explanation could be that the considered VASP is part of a larger company structure and that the company engages next to VASP activities also in non-VASP-related business activities. In that case, the reported balance sheet items might contain aggregated business activities, whereby it is difficult to disentangle the specific positions related to the crypto activities of the \VASPC. As a result, the proxy variable from the balance sheet might then overestimate the actual figure we are interested in.

Furthermore, this VASP operates with multiple DLTs and also exchanges stablecoins, but the cryptoasset wallets we analyzed do not hold any USDC, USDT, or DAI.

\subsubsection{\VASPD} 

\begin{figure}[h]
	\centering
	\includegraphics[width=\textwidth]{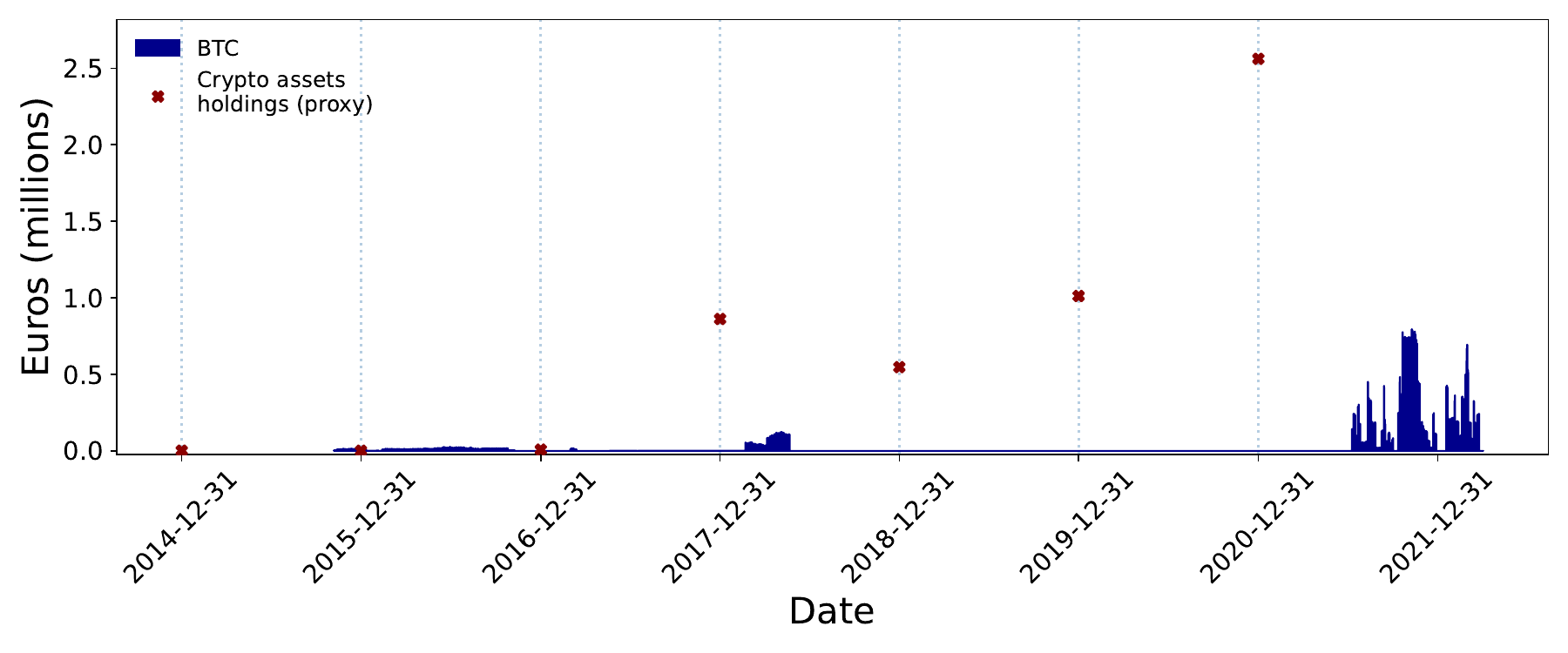}
	\caption{\textit{\textbf{Estimation of the bitcoin holdings in Euro of \VASPD.} On-chain and off-chain data are comparable only in 2015 and 2016.}} 
	\label{fig:VASP_D_BTC}
\end{figure}

\paragraph{Observations}
\VASPD is the last we analyze; values are shown in Figure~\ref{fig:VASP_D_BTC}. This VASP bases its services on the purchase and sale of bitcoins. For this VASP, using both attribution tags in the TagPack database mentioned above and re-identification strategies, we could only gather information for a few months in between 2014 and 2017 and after 2021. The results are consistent only for the years 2015 and 2016, when the VASP held very small amounts of cryptoassets, if compared to the subsequent years.

\paragraph{Findings}

Similarly to \VASPC, we could not collect sufficient data to obtain comparable values to the figures reported in the balance sheets. As for \VASPB, the Bitcoin addresses we gathered through manual transactions identify clusters whose transaction history only dates back to a few months (mid-2021). Again, this highlights that re-identification is less effective for Bitcoin than Ethereum addresses.

The data gap between 2018 and 2020 reveals another issue: likely, after 2017, funds were moved to other addresses that are not reused with those in our sample. VASPs apply different strategies to organize their cryptoasset transfers and holdings, e.g., to create new addresses for each transaction, or reuse them. If they are not reused, cryptoasset holdings can be held at multiple apparently unrelated clusters that can change over time.

%% file: tables/customers.tex
\begin{tabular*}{\textwidth}{@{\extracolsep{\fill}}lccccccc}
\toprule
 & Austria & EU$^{(*)}$  & Offshore & Embargo & Grey list & Other & Total \\
\midrule
Natural persons  &  \num{326660} & \num{1279132}  & \num{1160}  &  \num{1183} &  \num{36421}  &  \num{141491} & \num{1785747} \\
Legal persons  &  \num{326} & \num{147}   & \num{2}  & -  & -  & \num{26} & \num{501}  \\ 
\bottomrule
\multicolumn{8}{l}{\footnotesize $^{(*)}$excluding Austrian customers}\\
\end{tabular*}

%% file: sections_original/5_1_data_gap.tex

\section{Closing The Data Gap}
\label{sec:discussion}

We presented an approach to measure the cryptoasset holdings of VASPs by correlating data from multiple on-chain and off-chain sources. Empirical analysis of four VASPs reveals that only two of them show consistent comparisons of on-chain and off-chain data, indicating potential data-related problems. In this section, we systematically discuss the encountered data issues and provide suggestions for possible improvements. 

\subsection{On-chain data issues}
 
\paragraph{Different wallet management strategies}

VASPs employ diverse approaches to manage their cryptoasset transfers and holdings. While some create new addresses for each user transaction, others might reuse addresses. Moreover, their approach varies when dealing with UTXO-based or account-based ledgers. We observed that VASPs deploy user-specific Ethereum smart contracts wallets for each customer and subsequently forward the funds to a collector wallet. We did not observe this pattern with Bitcoin. This organization strategy makes it more challenging to identify cryptoasset holdings associated with VASPs. Identification largely relies on heuristics approaches, which can produce false positives and are often inadequately understood.

\paragraph{Lack of attribution data}

Another issue concerns the lack of attribution data, i.e., associations of addresses with additional contextual information allowing the identification of their owner. Our attribution dataset contains more than \num{265000000} deanonymized Bitcoin addresses and \num{278244} tagged Ethereum addresses.
Furthermore, we have conducted additional manual transactions with their services to identify and tag the specific addresses associated with the VASPs investigated in our study. Despite this, the resulting data only provide a partial view of their holdings, as shown in the previous section.

Another issue associated with manual tagging is that it misses historical data. As we showed in Section~\ref{sec:data_methods} for VASP-12 and VASP-5, we could only trace the Bitcoin transaction history of a VASP back in time for a few months when using re-identification techniques.

\paragraph{Missing cross-ledger perspective}

The data collected for both ledger types face a common issue --- they may only represent a portion of the total cryptoassets holdings. This could be because manual transactions used to tag hot wallets, which are addresses used for daily deposit and withdrawal operations, may not successfully identify cold wallets, i.e., the addresses that manage most of the VASP funds. The latter are subject to stricter security measures that may prevent association with hot wallet addresses. Additionally, wallets such as VASP-2 and VASP-12 contain more funds than reported to authorities, making it difficult to differentiate between customers' funds and other cryptoassets managed under the same wallet, such as equity or private funds.

\subsection{Off-chain data collection issues}

In addition to on-chain data, we used all data sources currently available for VASPs in Austria: balance sheet data from the commercial register and data from the supervisory entities.

\paragraph{Long reporting periods} Balance sheets are only published yearly, and asset holdings might differ before and after the exact reporting due date. Thus, the balance sheet statements of VASPs are only partially suitable for assessing their solvency.

\paragraph{Missing breakdown by cryptoasset type}

Nevertheless, it is important to outline the type of data and a good reporting practice for such data to improve the transparency of virtual asset-providing companies. Not all firms report balance-sheet items for crypto and fiat asset holdings separately. In the data comparison in Section~\ref{sec:data_methods}, we sometimes needed to use proxies for some VASPs that overestimate the actual cryptoasset holdings of a VASP, primarily due to the aggregation of multiple items within the same balance sheet entry. It is, therefore, essential that VASPs report their fiat and crypto asset and liability positions at a reasonable frequency separately from other activities within a company's holding structure.

\paragraph{Subsidiary companies and different jurisdictions}

VASPs may be subsidiaries of larger corporations. For VASP-9, we could not precisely determine the proportion of assets attributable to the subsidiary we examined. Moreover, many companies operate in several countries and fall under multiple jurisdictions, which adds another layer of complexity.

\subsection{Limitations of our approach}

Other limitations related to our approach stand out. First, data are extracted from the two major DLTs, Bitcoin and Ethereum, and only on a limited number of tokens supported by the latter. While these are the most relevant for market capitalization, including other DLTs and Ethereum tokens would be a straightforward improvement. Second, we gather Ethereum data by querying the account balances. Thus, we do not reconstruct balances from transactions, and we repeat the procedure on an interval of 10,000 blocks. We favor the approach based on querying the account states as it facilitates reproducibility at the cost of a lower granularity. We also note that this time interval can be easily changed with a shorter one. Third, our current approach is limited to end-year of 2021, but the analysis can potentially be extended to subsequent years.

\subsection{Towards a systematic assessment of proof of solvency}

Having discussed the data issues and limitations of our approach, we would like to sketch out our vision for a more systematic, reliable, and highly automated assessment of proof of solvency. 

\paragraph{Assessing proof of solvency today}

Fiat assets and liabilities are held at traditional financial intermediaries and undergo audits based on established standards. 
On the other hand, cryptoassets are held in cryptoasset wallets, scattered across various, potentially privacy-preserving DLTs, and are not subject to systematic and consistent audits. By measuring the cryptoassets held by one VASP, we can validate the amounts reported in the balance sheets. Given that the difference between assets and liabilities on a balance sheet equals equity, our method offers an initial, systematic validation and proof of solvency.
However, balance sheets currently disclose crypto and fiat deposits from customers under one balance sheet position. Thus, we cannot answer whether the VASPs retain the customer funds in crypto or convert them to fiat (or vice-versa). 

\paragraph{Improving on-chain data reporting}

Regarding on-chain data, we note that determining the solvency of VASPs is unfeasible without knowledge of the crypto addresses they control. Hence, any auditing entity must be aware of the on-chain cryptoasset holdings a particular VASP manages.
Furthermore, sharing a list of on-chain wallet addresses alone is insufficient. In a system with weak identities, anyone could hold the corresponding private keys and control the associated funds. VASPs need to prove they also control the funds they hold in custody for their users. 

Revealing a list of on-chain wallet addresses and transferring funds proves that a VASP possesses and manages specific funds. However, this approach can create privacy, security, and operational efficiency concerns. One way to mitigate these issues is to share this information only with trusted entities such as certified auditors or regulatory authorities. Furthermore, this approach would not disclose any information on actual user deposits.

Finally, in addition to disclosing their on-chain wallets, VASPs should provide additional metadata describing the use of these wallets. Most importantly, they should differentiate between hot and cold wallets and customer and non-customer (corporate) wallets. With hot wallets, they could also distinguish between deposit and withdrawal wallets and specify whether they are used per customer or across customers. In addition to the amounts contained therein, it would also be important for auditors to know what digital and physical security measures are taken to prevent cold wallets from being compromised.

\paragraph{Improving off-chain data reporting}

On the off-chain side, reporting requirements for a VASP should include a breakdown of asset holdings differentiating between fiat and crypto holdings. Such a breakdown is necessary for items on the asset side but also for items on the liability side. A step towards even more granularity is to differentiate the crypto items according to major cryptoassets and to provide wallet information on the storage of crypto asset holdings and liabilities. To understand the implications of VASPs on financial stability, frequent and detailed reports on the distribution of who are the counter-parties of VASPs (private customers, companies, other VASPs, \dots) and concepts of how and where crypto assets are stored are necessary information.

\paragraph{Enhancing VASP solvency assessment} 

One possible strategy to improve the assessment process is to use cryptographic primitives.
The academic literature has already proposed cryptographically secure proof-of-concept implementations for proofing the solvency of cryptoasset exchanges. \citealp{decker2015making}, in particular, proposed an audit process in a trusted computing environment that exploits digital signatures on their associated addresses for proofing reserves. Merkle trees, instead, are used to prove the total size of user deposits without directly leaking user-specific information.
This technique has already been implemented by several centralized exchanges (e.g., Binance\footnote{\url{https://www.binance.com/en/proof-of-reserves}}). However, that method has two flaws: first, an attacker that controls many accounts could still potentially learn a significant amount about the exchange's users; second, Merkle trees could allow an exchange that has more customer deposit assets than reserves to make up the difference by adding fake accounts with negative balances. To improve the privacy and robustness of that approach, \citet{buterin2022cexs} recently proposed to use ZK-SNARK to prove that all balances in the tree are non-negative.

A more forward-thinking strategy goes in the direction of automation. Given access to both on- and off-chain data with specific detail and granularity, the entire audit process could be streamlined and performed more systematically, frequently, and reliably than current methods allow. In line with this perspective, \cite{auer2019embedded} introduced the concept of ``embedded supervision'' enabling automated monitoring of decentralized finance (DeFi) services to ensure compliance with regulatory objectives. 
\cite{buterin2023blockchain} studied an automated privacy-enhancing protocol that utilizes smart contracts and ZK-SNARKS to prove that the users' assets were received from lawful sources. 
Additionally, \cite{eichengreen2023stablecoin} suggest that real-time audits carried out by independent proof-of-reserve systems and facilitated by smart contracts could effectively mitigate the threat of stablecoin devaluation. 

In conclusion, it is noteworthy that, according to Article 29 (1) of the Austrian AML-Act, the FMA already possesses the authority and legal mandate to request essential data from all obliged entities (i.e., VASPs) at any time on all issues that are addressed in the Austrian AML-Act and Regulation (EU) 2015/847, e.g. a list of cryptoasset addresses under their control\footnote{\url{https://www.ris.bka.gv.at/eli/bgbl/i/2016/118/P29/NOR40189690}, \url{https://eur-lex.europa.eu/eli/reg/2015/847/oj}}.

%% file: sections_original/6_conclusions.tex

\section{Conclusions}
\label{sec:conclusions}

In this work, we investigate 24 VASPs registered with the Austrian Financial Market Authority (FMA) at the end of 2022. We aim to provide an empirical approach to assess their solvency status, by measuring their cryptoasset holdings across time and distributed ledgers. To do so, we cross-reference data from three distinct sources: publicly auditable cryptoasset wallets, balance sheet data from the commercial register, and information from supervisory entities. We begin by describing the financial services they offer, the virtual assets they support, and compare them to conventional financial intermediaries. Their core financial activity can be compared to money exchanges, brokers, and funds, rather than to commercial banks.
Furthermore, we provide regulatory data insights showing that their yearly incoming and outgoing transaction volume in 2022 amounted to 2 billion EUR for around 1.8 million users.

Next, we implement address clustering algorithms and entity identification techniques to reconstruct their cryptoasset flows on the Bitcoin and Ethereum blockchains and compare their net positions to balance sheet data from the commercial register. We focus on four VASPs for which we could gather information both on their cryptoasset transactions and balance sheets. These four entities cover around 99\% of the Austrian VASP transaction volumes measured in total assets. With our approach, we find proof, for two VASPs out of four, that they control enough assets to fulfill liabilities and obligations against customers, i.e., they meet the capital requirements, while we could not collect enough data for the remaining two.

Then we discuss the data collection-related issues and suggest solutions towards better assessing a VASP solvency. In particular, we remark that any entity in charge of auditing requires proof that a VASP actually controls the funds associated with its on-chain wallets. It is also important that a VASP reports fiat and crypto asset and liability positions, broken down by asset types at a reasonable frequency.

In conclusion, our approach highlights the need to address the identified data gaps in the current data collection process and provides a starting point for developing more effective strategies to systematically assess the solvency status of virtual asset service providers.

\section*{Declaration of Competing Interest}

The authors declare that they have no known competing financial interests or personal relationships that could have appeared to influence the work reported in this paper.

%% file: sections_original/supplemental_material.tex

\section{Supplemental material}
\label{sec:supplemental}

\setcounter{figure}{0}
\setcounter{table}{0}

\begin{figure}[hp]
	\centering
	\begin{subfigure}{.5\textwidth}
		\centering
		\includegraphics[width=\linewidth]{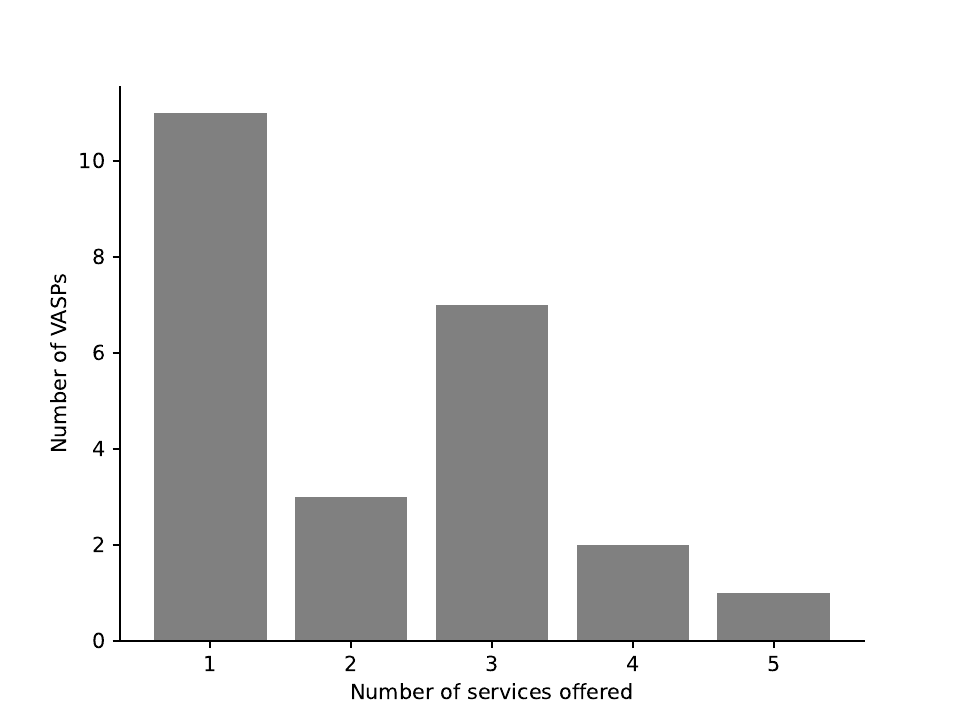}
		\caption{}
		\label{fig:1}
	\end{subfigure}%
	\begin{subfigure}{.5\textwidth}
		\centering
		\includegraphics[width=\linewidth]{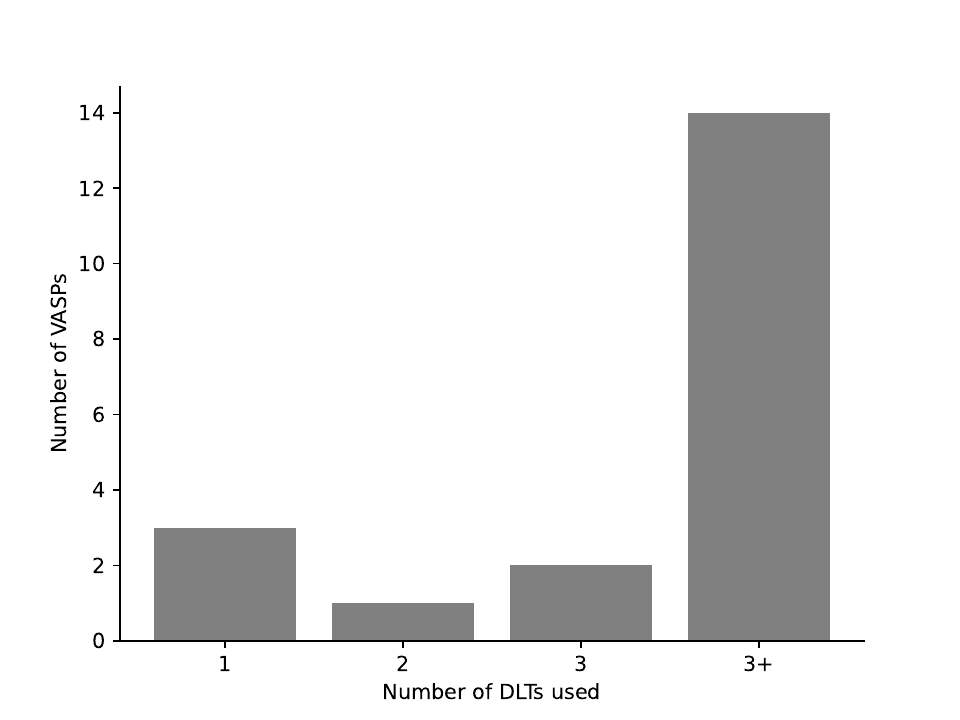}
		\caption{}
		\label{fig:2}
	\end{subfigure}
	\caption{\textit{\textbf{The Austrian VASP landscape}. (\subref{fig:1}) histogram showing how many services VASPs offer; (\subref{fig:2}) histogram showing how many DLTs VASPs utilize.}}
		\label{fig:VASPs_2}
\end{figure}

In this appendix we report additional information about how many VASPs offer several services or support multiple coins (Figure~\ref{fig:VASPs_2}). Most VASPs ($N = 14$) provide services on cryptoassets of three or more different DLTs; $N = 11$ VASPs provide only one service, and $N = 7$ offer three different services.

Furthermore, we provide further insights on the address clustering technique utilized and on the data gathering procedure. 
The address clustering we implement relies on the multi-input technique discussed in~\citet{androulaki2013evaluating,ron2013quantitative,Meiklejohn2016}. It assumes that the addresses utilized as input in a Bitcoin transaction must be controlled by the same entity. If addresses are reused across transactions, then multiple input addresses can be associated as belonging to the same entity.
We note that in this study we computed the VASP balances by analyzing the flows at the level of clusters. We also repeated the analysis at the address level, and found minor inconsistencies most likely due to rounding errors.

Next, we report additional information on the addresses we utilized to identify VASPs. The full list of addresses used is reported in Tables~\ref{tab:btc_addresses} and~\ref{tab:eth_addresses}. As explained in Section~\ref{sec:data}, we first exploited a collection of tagpacks that associate addresses to the entities controlling them.

Furthermore, to increase our dataset sample, we conducted manual transactions against the VASPs that were already in our sample. This led to the identification of 9 new addresses and their corresponding clusters, that we highlight in light gray in Tables~\ref{tab:btc_addresses} and~\ref{tab:eth_addresses}.

We further expanded the dataset by conducting transaction pattern analyses of the new collected addresses. We identified 7 additional addresses that follow specific patterns indicating the redirection of funds from temporary addresses to collector wallets. These are highlighted in Tables~\ref{tab:btc_addresses} and~\ref{tab:eth_addresses} in darker gray.

Finally, Table~\ref{tab:VASPs_expand} reports the VASP categorization by their service offering according to their online documentation.

\begin{table}[!h]
	\scriptsize
	\caption{List of Bitcoin cluster-defining addresses.}
	\label{tab:btc_addresses}
	\hspace*{-0.75cm}
	\input{tables/btc_addresses_used.tex}
\end{table}

\begin{table}[!h]
	\scriptsize
	\caption{List of Ethereum addresses.}
	\label{tab:eth_addresses}
	\hspace*{-0.75cm}
	\input{tables/eth_addresses_used.tex}
\end{table}

\begin{table}[h]
	\renewcommand{\arraystretch}{1.3}
	\centering
	\footnotesize
	\caption{VASPs categorization by their service offering --- VASP-specific observations.}
	\label{tab:VASPs_expand}
	\input{tables/VASPs_full.tex}
\end{table}

%% file: tables/btc_addresses_used.tex
\begin{tabular}{llll}
\toprule
Address & Label & Address & Label \\
\midrule
372iojpPqRP2e3oh7eUs1VZowS8kDkMrff & VASP A & bc1qagw7nw7na3ev9yln2yeggp2wd26h6lxm25um3v & VASP A \\
bc1pgwsxt7ww3s2rsa8g0jpjyyvyd30xm4kg3skhm5... & VASP A & 3A4p29nPvfFGBa7a1KNFsioRyKwv9GdvBL & VASP A \\
35TviLjv9zD91Q9N7X3kcqstZdBusTpqNe & VASP A & bc1q3nmcqejgyldax0ekfcj2w5zcjgmf43wwdfcugj... & VASP A \\
36iUjkZwQAvkz8StPqudfdQnkq57e5tZTL & VASP A & 3BqHzZAgqEE8pYz2cYikWrvwW8UZSL4Y6w & VASP A \\
3AQGNPTTY4D3a5AfG2C3ktQdkBfbJzFaZG & VASP A & 13b84xNArAK8T4KbjbqaHDVREVC11CaAqQ & VASP A \\
bc1qwqgtpvh7u5fcf043y7ze9934jct8g74ducmpat & VASP A & 3F8z98fknEknhwVwmfDLo9kwdzrFo5PGZe & VASP A \\
3Ms6oxUU9vdek3tFS2UxtC6v77aLurfTnV & VASP A & 18p9Ftp3m4435tdpZTvoBsm3yjUgkvTF2b & VASP A \\
3QVRCiqw2HRCf3vnbkiQmNYSHcPnuVy8Lj & VASP A & 3AkxoL4gEUvZLjAikYn49J7cLDWkYsZyqy & VASP A \\
3Djd57FoRZjjT5rJx6WWmibgA2esSSEPLs & VASP A & 38p43PUvNEXLZMyHzCddbfznzU44RPokc4 & VASP A \\
3AqPuD4ZdHPjnN79mKNCk46mpMMmyFPgQd & VASP A & bc1q908vpavz7f4aej5aw7d6qku5u3llrcfc42d680 & VASP A \\
3AZfcZ8NqaGfB5PHGCpQ29QXE8TgbmyxGL & VASP A & 3BNHfkoeJ5hTAtGviQKGtuHdDnUP4fS67s & VASP A \\
\cellcolor{gray!15}bc1qufd8s2e7l29mkpnxh29z4zmyhqhz4jh8u62j64 & VASP A & bc1qz0klefh90jpnk48krl8kka22z034g9dv9mfwxy & VASP A \\
bc1qla7dpxf23yqghxvgsnm6e46e4kgvgk3ghwanhs & VASP A & 3LLxkyXbymE6gyncUtWCtDm2kMBYnh8Y5H & VASP A \\
36y5oizZAKNSXEEcvQnHHjBPzgDoyUgH6U & VASP A & 3NjTTuWoAimdkF6kU29Wg3TaX9Y41xJk2Q & VASP A \\
3EwXtQN38N7bczhtssTaQXQ8V1ESzGCWXp & VASP A & 36M14LUZ5QdCpPAvRhbyuPrbySCixana7q & VASP A \\
35WA9LqtgrFtnW99ZjG8SigwsDJfvCFNN6 & VASP A & 3NeVui16zjEFB5DAkZYnEzFzemwffxdLGC & VASP A \\
3FJuyixsUGR8aCHaJXotyu5mFTdgQTUGN8 & VASP A & 365dwi6ePGn5NQwrJPgpkn4AZtY7U1DAzd & VASP A \\
bc1qlvaaw7qg60g4r34uqvcykcyeprqq3gzunxl959 & VASP A & 1G2k682CGjbgDt5TA9LPvUUZALVnGRJNbX & VASP A \\
32oA2zy5g8bsH6ibA84zyTiot57UNn6cZk & VASP A & 3ABGvFN2KUmyqCciVqGRWP7qozXC3GJY1t & VASP A \\
38Hed9w9ipv7chiztoXmVFpE9QoKK4A5e2 & VASP A & 37sF88thJvtaHKtDxWam1iNE967VXMYmZ4 & VASP A \\
37h3xPkADPXz2Gf13LCbcQGPWSs9VUbijD & VASP A & 1hseBvYYoSmrwLRxAiW8mZJ4PuDCk8vC6 & VASP A \\
3BSLDuJ5BniD89ss7AixMAaxzNLxz73tqC & VASP A & bc1qpm8uuck4nd2drmah7pg4wn2ry8svpkzc5g9h2d & VASP A \\
3P51Su6oAUgKnqHA8RwAB17iXtNh1MqrLN & VASP A & 12j9hDEUjjPx9r5fP6S6QwFrDt86pipHY5 & VASP A \\
bc1q90ln97x30cu07k0c6f4sd5d0w304csx7wl65af & VASP A & 3BaZvbaJMoLw7xDjk5kuujMCRkebSh35x7 & VASP A \\
3GXL7pb7AQU2PuDUx2FanShaSxfaijSuoq & VASP A & 3EX9Bufgb7E5dy63vwSPfBfbvuxz6AEdJ8 & VASP A \\
3NSvc7wp661GKovmtsUowXirVsPZaaXfEA & VASP A & 3PGzrQkvYshnDAdQb8mcajcXT69SLAUqNM & VASP A \\
3NKeW5Zp1C6MTVDvu45FrPwv64wmgwokMx & VASP A & 3KbJUSyQxEQi7fgie2U8oEbEMPgzEXPhxb & VASP A \\
1DQUovque483G1qogond6ar2jekVbBbQoa & VASP A & 3H1M9CJaGTmNWiKiGJSdjqvYK4jvgCkWZD & VASP A \\
3Q3R7ohNcmPAP8cJBxbGHvJ11dtD4AiWA4 & VASP A & 3R2Gf3pLjh4T45VoJbpXNUErjyu7iNi9Lc & VASP A \\
3JcHEnvorB1iYjb3PtNGGWZRekE3hrYZg2 & VASP A & 3LWD798xiNsHEj3tHmseTvMGicTyMw89yH & VASP A \\
3PZVQhKmXVyeF1u58GTpshWwba6MrVwoxx & VASP A & bc1qyw3t7aapdgfc6nnsq9cjt9snyyzm0h2m5cvz6c & VASP A \\
1AmicLpEygMr6XbifV3v89HEJE1JAob6MP & VASP A & 3CFSNanniaS3TgXvKkDQxSdjaqFiZhYFjo & VASP A \\
bc1q2fuj4pdlvftes962udrl6rg8hserg5al9mwwq2 & VASP A & 3GiJn8VRRfuCWLcepv2vpz3SijnWDZbJNV & VASP A \\
1AXhjxmb84UXUuKMz1kiotrEWkYGHdr2pk & VASP A & 3QTejSeVyUEyi1bdcYLBrbaW2Y7PcT1mYC & VASP A \\
33CcPBjiX8BSoAq1bnSwBykNvDu6D2ikPe & VASP A & 1J8p8e8XerfNsT2rHPsT4EGXewcKN4TcZC & VASP A \\
3PoVeBNfNhyCkAXfY7zXFvevqxvAyjkZHo & VASP A & 139JQeoAHTUvHbhZQrumMfvTVjjj4XHWqJ & VASP A \\
1EnxErtRRpfshfZHCGj2dhVfKqNUvqB6uV & VASP A & 349o5hFXzajRqd5keuhc4Vpvk5UnJvKDwe & VASP A \\
3QDrBALoR558xr7AQ2qbUweLYMdkhTzGNi & VASP A & \cellcolor{gray!15}3EBhkMekkAbZnjZKj278S9Ep5CBopUwJAJ & VASP B \\
\cellcolor{gray!40}bc1qa2sshk7mf8ln6dcz3yshga4ry6yad40fafks3m & VASP B & \cellcolor{gray!15}bc1q5upx09sj98mkql9uchnjf2fz9zk0e5cp28393n & VASP B \\
\cellcolor{gray!15}bc1qekd92htscmj9jhnhe4c8uw5acnvgsl6pphzumg & VASP B & 36WHTtkZ5jmwNk8ZyEQK7DYyjv99REjiXg & VASP C \\
3CoiP8UBMLkbqCftfE9VWfpTwFdTk4v8tk & VASP C & \cellcolor{gray!15}3Gc1VcYkVZN7onTxxMkRLqtmokzxpaFHGG & VASP C \\
\cellcolor{gray!40}bc1qnxx404l623aaejv252htxh39s42te6wu6aa8ay & VASP D & 3Ai4XQYJyD3NRToE3PA8odv6SLqGLgCaaC & VASP D \\
\cellcolor{gray!15}bc1qp7h0wh3hxyax5nxawv36vxvpl7gml6t7dqr2tp & VASP D & 3BoECgzNsS5NqNHn5wvtR9xVmyxCFDcGuD & VASP D \\
33cXNWciLE74bnRjk6Dz2k1fGVSxByewc8 & VASP D & \cellcolor{gray!15}bc1qlu2qrxcxzkd2jkgzpj7ldspn8myfraxgwkyg2h & VASP D \\
\bottomrule
\end{tabular}

%% file: tables/eth_addresses_used.tex
\begin{tabular}{llll}
\toprule
Address & Label & Address & Label \\
\midrule
\cellcolor{gray!40}0x1eDB8A5d51880c81bA6B4812485c3dC16085fC39 & VASP A & 0x2754b28227F041a66c46509D5620782BFC4766EF & VASP A \\
0x74dEc05E5b894b0EfEc69Cdf6316971802A2F9a1 & VASP A & \cellcolor{gray!40}0xCC6E3Fd35F034F5baA27b6E74DCB197f084A8721 & VASP A \\
\cellcolor{gray!40}0xF32682d5F99ba4143532618d6f516859a055Ea06 & VASP A & \cellcolor{gray!15}0xDd0b0DE8D457b6FC20e8f9E9dd5a38A525EF4258 & VASP B \\
\cellcolor{gray!40}0x0067F95A79c3C404a9d128168DDFDf3cB70c0852 & VASP B & \cellcolor{gray!15}0x16076b17bd55a2ebbda011d39dca8916094a0c38 & VASP C \\
\cellcolor{gray!40}0xC3b7336D5A5158215599572012CeDd4403A81629 & VASP C &   &   \\
\bottomrule
\end{tabular}

%% file: tables/VASPs_full.tex
\begin{tabular*}{\columnwidth}{@{\extracolsep{\fill}}lccccc}
	\toprule
	Name & \makecell{Custody \\ services} &  \makecell{Buy/Sell \\ services} & \makecell{Payment \\ processing} &  \makecell{Consulting \\ services} & \makecell{Trading \\ platform} \\ 
	\midrule  \vspace{0.2em}
	VASP-0  &  N  &  Y  &  N  &  N  &  N \\ \vspace{0.2em}
	VASP-1   &  N  &  Y  &  N  &  N & N \\ \vspace{0.2em}
	VASP-2    &  Y  &  Y  &  N  &  N  &  Y   \\ \vspace{0.2em}
	VASP-3  &  Y  &  Y  &  N  &  Y  &  Y \\ \vspace{0.2em}
	VASP-4  &  Y  &  Y  &  N  &  Y  &  N  \\ \vspace{0.2em}
	VASP-5	&  Y  &  Y  &  N  &  N  &  N  \\ \vspace{0.2em}
	VASP-6 & Y  &  Y  &  N  &  Y  &  N \\ \vspace{0.2em}
	VASP-7 & Y  &  Y  &  N  &  Y  &  N \\ \vspace{0.2em}
	VASP-8 &  N  &  Y  &  N  &  N  &  N  \\ \vspace{0.2em}
	VASP-9 &  Y  &  Y  &  N  &  N  &  N  \\ \vspace{0.2em}
	VASP-10  &  N  &  Y  &  N  &  N  &  N \\ \vspace{0.2em}
	VASP-11&  N  &  Y  &  N  &  N  &  N   \\ \vspace{0.2em}
	VASP-12  &  N  &  Y  &  N  &  N  &  N \\ \vspace{0.2em}
	VASP-13 & Y  &  Y  &  N  &  Y  &  N  \\ \vspace{0.2em}
	VASP-14	&  N  &  N  &  Y  &  N  &  N \\ \vspace{0.2em}
	VASP-15 & Y  &  Y  &  N  &  Y  &  N  \\ \vspace{0.2em}
	VASP-16  &  Y  &  Y  &  N  &  N  &  Y  \\ \vspace{0.2em}
	VASP-17 &  Y  &  Y  &  N  &  Y  &  N \\ \vspace{0.2em}
	VASP-18  &  N  &  Y  &  N  &  N  &  N  \\ \vspace{0.2em}
	VASP-19 &  Y  &  N  &  N  &  N  &  N  \\ \vspace{0.2em}
	VASP-20  &  Y  &  Y  &  Y  &  N  &  N \\
	\bottomrule
\end{tabular*}